\documentclass[aps,prl,twocolumn,showpacs,showkeys,preprintnumbers,superscriptaddress,amsmath,amssymb]{revtex4-1}
\usepackage[T1]{fontenc}
\usepackage[utf8]{inputenc}

\usepackage[dvipsnames,usenames]{xcolor}

\usepackage{amsmath}
\usepackage{amssymb}
\usepackage{xcolor}
\usepackage{wrapfig}
\usepackage{lipsum,graphicx}
\usepackage{float}
\usepackage{gensymb}
\usepackage{sidecap}
\usepackage{textcomp}
\usepackage{hyperref}
\usepackage{hyperref}
\usepackage{times}

\DeclareMathOperator\erfc{erfc}

\def\sx2{\sin{\frac{k_x}{2}}}
\def\sy2{\sin{\frac{k_y}{2}}}
\def\cx2{\cos{\frac{k_x}{2}}}
\def\cy2{\cos{\frac{k_y}{2}}}

\def\sqsx2{\sin^2{\frac{k_x}{2}}}
\def\sqsy2{\sin^2{\frac{k_y}{2}}}
\def\sqcx2{\cos^2{\frac{k_x}{2}}}
\def\sqcy2{\cos^2{\frac{k_y}{2}}}

\begin{document}

\title{Photoinduced renormalization of Dirac states in BaNiS$_2$} 

\author{N. Nilforoushan}
\affiliation{Laboratoire de Physique des Solides, CNRS, Univ. Paris-Sud, Universit\'e Paris-Saclay, 91405 Orsay Cedex, France}
\author{M. Casula}
\affiliation{IMPMC , Sorbonne Universit\'e, CNRS , MNHN , 4 place Jussieu , 75252 Paris, France}
\author{M. Caputo}
\affiliation{Laboratoire de Physique des Solides, CNRS, Univ. Paris-Sud, Universit\'e Paris-Saclay, 91405 Orsay Cedex, France}
\author{E. Papalazarou}
\affiliation{Laboratoire de Physique des Solides, CNRS, Univ. Paris-Sud, Universit\'e Paris-Saclay, 91405 Orsay Cedex, France}
\author{J. Caillaux}
\affiliation{Laboratoire de Physique des Solides, CNRS, Univ. Paris-Sud, Universit\'e Paris-Saclay, 91405 Orsay Cedex, France}
\author{Z. Cheng}
\affiliation{Laboratoire de Physique des Solides, CNRS, Univ. Paris-Sud, Universit\'e Paris-Saclay, 91405 Orsay Cedex, France}
\author{L. Perfetti}
\affiliation{Laboratoire des Solides Irradi\'{e}s, Ecole Polytechnique, CNRS, CEA, 91128 Palaiseau cedex, France}
\author{A. Amaricci}
\affiliation{CNR-IOM DEMOCRITOS, Istituto Officina dei Materiali, Consiglio Nazionale delle Ricerche, Via Bonomea 265, I-34136 Trieste, Italy}
\author{D. Santos-Cottin}
\affiliation{IMPMC , Sorbonne Universit\'e, CNRS , MNHN , 4 place Jussieu , 75252 Paris, France}
\author{Y. Klein}
\affiliation{IMPMC , Sorbonne Universit\'e, CNRS , MNHN , 4 place Jussieu , 75252 Paris, France}
\author{A. Gauzzi}
\affiliation{IMPMC , Sorbonne Universit\'e, CNRS , MNHN , 4 place Jussieu , 75252 Paris, France}
\author{M. Marsi}
\affiliation{Laboratoire de Physique des Solides, CNRS, Univ. Paris-Sud, Universit\'e Paris-Saclay, 91405 Orsay Cedex, France}

\date{\today}
\begin{abstract}
By means of pump-probe time- and angle-resolved photoelectron spectroscopy, we provide evidence of a sizeable reduction of the Fermi velocity of out-of-equilibrium Dirac bands in the quasi-two-dimensional semimetal BaNiS$_2$. First-principle calculations indicate that this band renormalization is ascribed to a change in non-local electron correlations driven by a photo-induced enhancement of screening properties. This effect is accompanied by a slowing down of the Dirac fermions and by a non-rigid shift of the bands at the center of the Brillouin zone. This result suggests that other similar electronic structure renormalizations may be photoinduced in other materials in presence of strong non-local correlations. 
\end{abstract}
\pacs{}

\maketitle

The recent discovery of Dirac and Weyl fermions in crystalline solids \cite{Armitage2018,Liu2014a,Xu2015,Wang2012,Borisenko2014} provides a privileged experimental environment for the study of these exotic electronic states. The unique properties of these states also suggest novel applications; for instance, the manipulation of Dirac fermions with ultrafast light pulses is attractive in view of novel concepts of optoelectronic devices combining unprecedented mobility and subpicosecond response, as well as of enabling photoinduced topological phase transitions \cite{Floquet2013, Weber2017, Sie2019}.
The electronic band structure of these materials is characterized by several linearly dispersing bands that originate from and are protected by the interplay between real- and momentum- space properties \cite{Young2012, Young2015,Tar12, Bahramy2017}. The ability to tailor the properties of these bands is a prerequisite for any functional application of relativistic Dirac fermions. 
 

In the present paper, we demonstrate that Dirac bands can be renormalized using ultrafast light pulses in the quasi two dimensional Dirac semi-metal BaNiS$_2$. Combining time-resolved ARPES (trARPES) and \emph{ab initio} calculations, we show that the key underlying mechanism is a change in the dynamical screening due to long-range correlations. The transient photoinduced renormalization is fully reversible and non-thermal in nature, as the effect of temperature on the massless Dirac particles is qualitatively different. 

The proposed scenario of photo-induced out-of-equilibrium electronic states is strongly corroborated by the experimental observation that the transient response of the Dirac bands is qualitatively different from that of other bands, consistent with our \emph{ab initio} calculations. Indeed, besides the massless Dirac fermions, the Fermi surface of BaNiS$_2$ also exhibit small electron pockets at $\Gamma$ \cite{SantosCottin2016}. Under the effect of ultrafast optical excitation, the energy levels at $\Gamma$ present non-rigid shifts depending of their orbital nature, in full agreement with our calculations. 

The ultrafast renormalization of the band structure has insofar been investigated only in insulating phases \cite{Brouet2013, Rameau2014, Mor2017, Roth2019} or in thin films where the substrate significantly affects the the screening properties of the system \cite{Ulstrup2016,Nicholson2018}. The present study on BaNiS$_2$  rather shows that the transient effect of ultrafast light pulses significantly alters the Fermi velocity of the relativistic Dirac states, and at the same time shifts the electron pocket at $\Gamma$ above the Fermi level. These results point at a dramatic change in the fermiology and in the transport properties of this semi-metal.  

\begin{figure}[H]
\centering 
\includegraphics[width=\columnwidth]{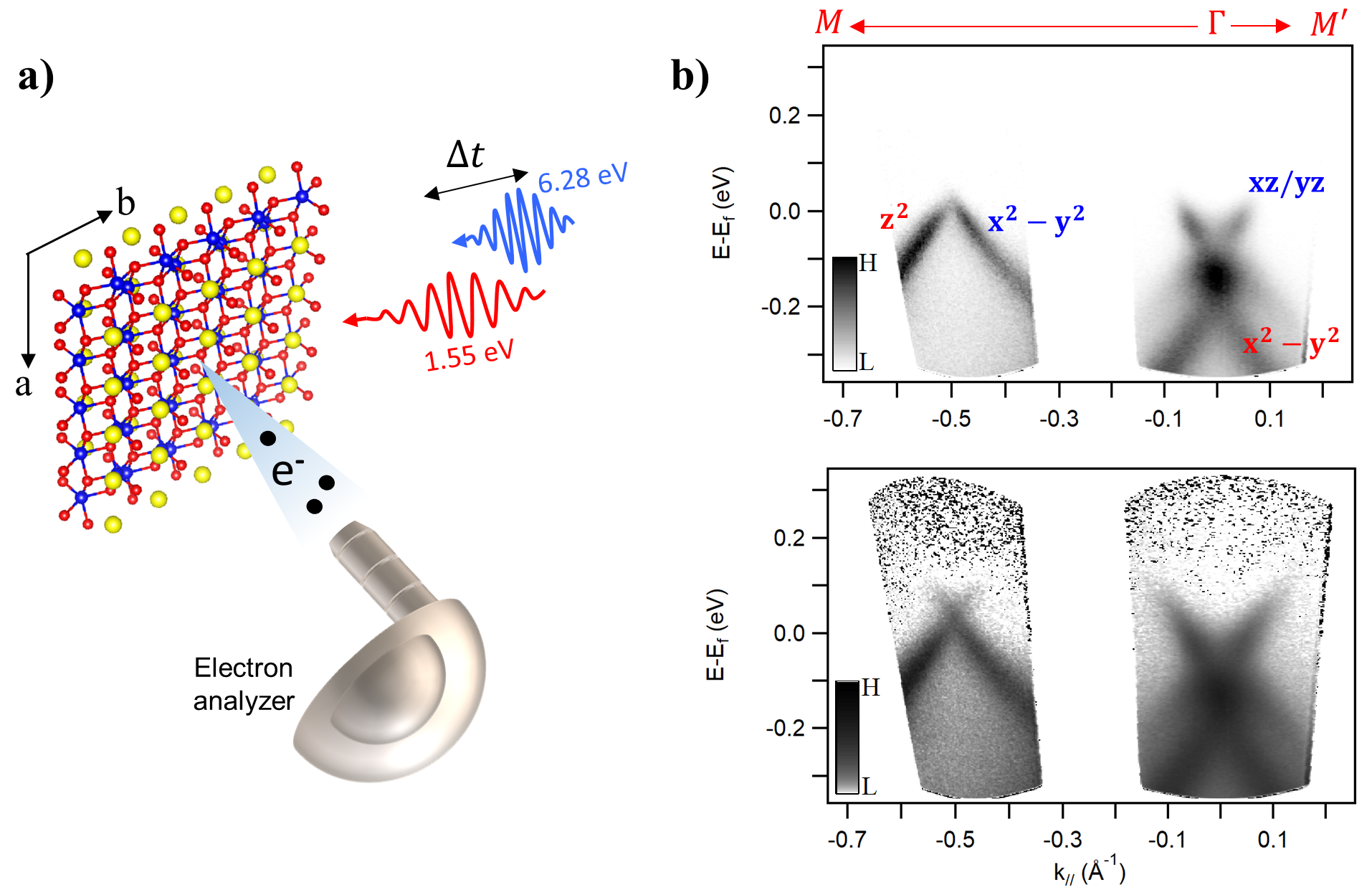}
\caption{
(a) Experimental geometry of the tr-ARPES setup. 
(b) Top: reference spectra of the bands along $\Gamma - M$. Bottom: photoexcitation of the electronic states at a 250 fs delay with a pump fluence of 0.2 mJ/cm$^2$.}
\label{fig:Figure1}
\end{figure}

\begin{figure*}[th]
\centering 
\includegraphics[scale=0.5]{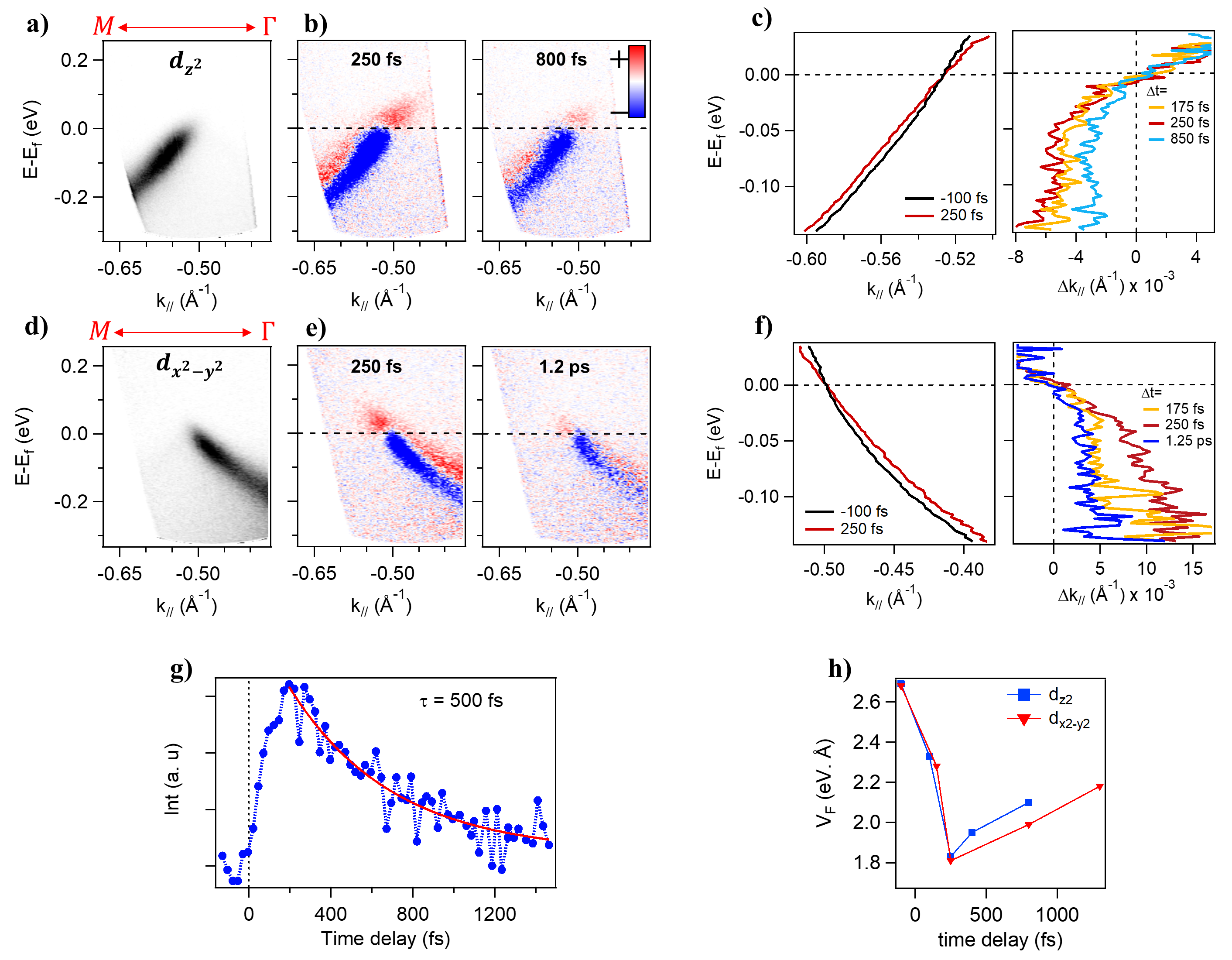}
\caption{
(a) ARPES reference image for the out-of-plane $d_{z^2}$ band. (b) Difference tr-ARPES images (after photoexcitation minus before photoexcitation): red and blue in the color scale indicate gain and loss of photoemission yield, respectively. 
(c) Band dispersion, $E(k)$, before and after arrival of the pump pulse, extracted as detailed in the text. The photoinduced band renormalization for different time delays is depicted on the right: the curves are obtained by subtracting the $E(k)$ curves at the negative delay from the $E(k)$ curves at positive delays. (d-f) same as (a-c), but for the $d_{x^2-y^2}$ band. In both cases, the photoinduced renormalization effect reaches a maximum around 250 fs. (g) Dynamics of the excited states photoemission yield, integrated above the Fermi level. The decay takes place with a time constant of 500 fs. (h) Dynamics of the Fermi velocity for the linearly dispersing bands obtained by integrating excited states above the Fermi level.}

\label{fig:Figure2}
\end{figure*}

For the present experiment, BaNiS$_2$ single crystals are cleaved in ultrahigh vacuum on the $a - b$ plane at the base temperature of 130 K. We drive BaNiS$_2$ out-of-equilibrium by applying 1.55 eV infrared (IR) pump pulses. Time-resolved snapshots of its evolution are taken with photoelectrons emitted with 6.28 eV ultraviolet (UV) pulses \cite{Faure2012}. The overall time resolution of our experiment is 80 fs. Fig.~\ref{fig:Figure1}(a) presents our time-resolved ARPES setup. The equilibrium band structure for different regions along the $\Gamma - M$ direction is obtained by acquiring ARPES spectra for different emission angles and is shown on the top panel of Fig.~\ref{fig:Figure1}(b). The $d$-orbital nature of each band, as deduced from previous \textit{ab-initio} calculations \cite{SantosCottin2016}, is also mentioned on the figure. The photoexcited band structure is shown on the bottom panel at a 250 fs time delay. 

\begin{figure*}[th]
\centering 
\includegraphics[scale=0.4]{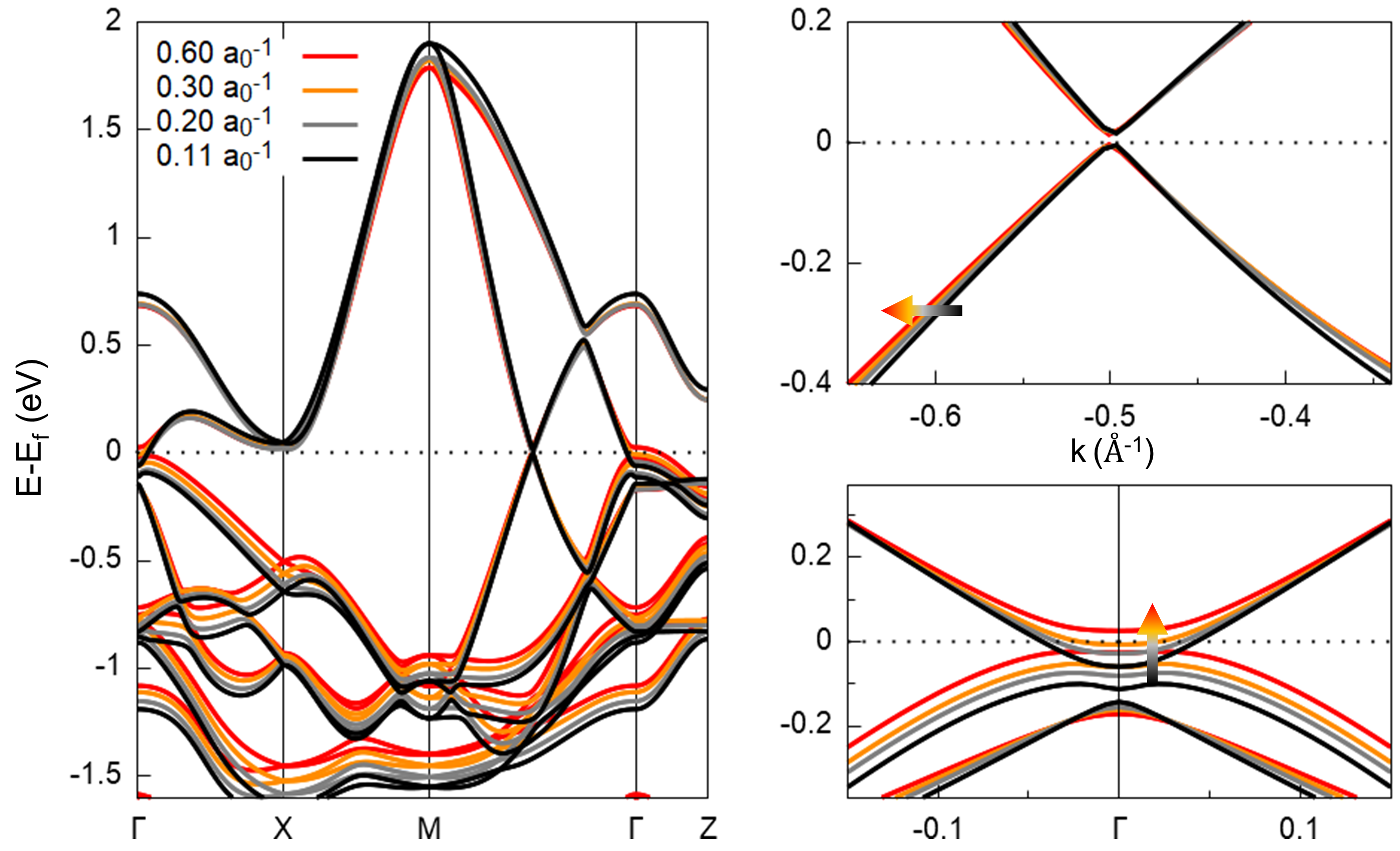}
\caption{(a) Band structure computed by the hybrid HSE06 DFT functional with modified exchange parameters. The exchange weight is $\alpha=0.07$, kept constant throughout the calculations. The non-local interaction range,  $\lambda=1/\omega$, is modified. Its original HSE06 value ($\omega=0.108$) corresponds to the system at equilibrium (black lines), while $\omega=0.6$  is representative of the photo-excited system prior to relaxation (red color). 
Relaxation is described by progressively decreasing $\omega$ from 0.6 to 0.108, corresponding to the reduction of screening due to electron thermalization towards equilibrium. (b) A zoom on the evolution of the Dirac states along the $\Gamma - M$ direction. (c) Behavior of the photo-excited band structure around the $\Gamma$ point, along $\Gamma - M$. The arrows depicts the effect of the increase of screening on the electronic bands.}
\label{fig:theory}
\end{figure*}

We first focus on the ultrafast dynamics of each Dirac band by tuning the light polarization as described elsewhere \cite{SantosCottin2016, arxivBNS}. 
Figs.~\ref{fig:Figure2}(a) and \ref{fig:Figure2}(d) show the reference spectra of the $d_{z^2}$ and $d_{x^2-y^2}$ bands, respectively. 
Figs.~\ref{fig:Figure2}(b) and \ref{fig:Figure2}(e) show a series of difference tr-ARPES spectra corresponding to the difference between images taken with positive and negative delays using a pump fluence of 0.2 mJ/cm$^2$. 
The difference spectra show an intensity gain (red) above the Fermi level due to the photoexcitation of electrons to the unoccupied states. As illustrated in Fig.~\ref{fig:Figure2}(g), these states reach a maximum of occupation around 250 fs due to electron scattering and impact ionization \cite{Caputo2018}. Subsequently, their population decrerases with a timescale of 500 fs. The relatively slow decay of hot carriers is ascribed to the limited phase-space available for scattering phenomena, a characteristic of semimetals \cite{Faure2013}. We now discuss the evolution of the valence bands below the Fermi level shown in Figs.~\ref{fig:Figure2}(b) and \ref{fig:Figure2}(e). These states display two distinct features: (i) an expected intensity loss (shown in blue) due to a transient depletion of the electron population: (ii) a remarkable intensity gain (shown in red), a signature of an ultrafast and time-dependent renormalization, \emph{i.e.} a non-rigid shift, of the Dirac states.
In order to study the dynamics of the band dispersion, $E(k,t)$, for each $\Delta t$ we fitted the momentum distribution curves (MDC's) of the ARPES yield, as reported in the supplemental material (SM). Figs.~\ref{fig:Figure2}(c) and \ref{fig:Figure2}(f) depict the band dispersion of the $d_{z^2}$ and $d_{x^2-y^2}$ band, respectively, before (-100 fs) and after (+250 fs) the arrival of the pump pulse. We clearly observe an energy-dependent band shift corresponding to a reduction of Fermi velocity. In the right panel of Figs.~\ref{fig:Figure2}(c) and \ref{fig:Figure2}(f), we plot the band renormalization, \emph{e.g.} the change in the wave vector at the given binding energy, for different time delays. Note that the non-rigid shift of the wave vector is larger as one moves away from the Fermi level. For example, $\Delta k$ even exceeds 0.01 \AA$^{-1}$ for the $d_{x^2-y^2}$ band. On the contrary, within our experimental resolution, the Fermi wave vector does not exhibit any shift and remains in its equilibrium position for all time delays. Fig.~\ref{fig:Figure2}(h) shows the dynamics of the Fermi velocity of both bands, where one notes a very large reduction up to 30\% for $t\sim 250$ fs. We would like to emphasize that at these time delays, the system is in a markedly non-thermal excited state (see Supplementary Material). To the best of our knowledge, such dramatic photoinduced renormalization of the band structure in a (semi-)metal, far from any electronic or structural phase transition, has not been reported before. 

In order to account for the above observation, we model the system by using \textit{ab initio} density functional theory (DFT). Previous BaNiS$_2$ calculations \cite{Klein2018a} showed that the low-energy physics of this quasi 2D system is driven by strong non-local correlations. Indeed, a satisfactory agreement between theory and quantum oscillation experiments has been achieved only by properly tuning the parameters of the screened exchange operator in the modified Heyd, Scuseria, and Ernzerhof (HSE) functional \cite{heyd2003hybrid,ge2006erratum,krukau2006influence}. The same agreement was not reachable within semilocal density functionals, such as the Perdew, Burke, and Ernzerhof (PBE) \cite{perdew1996generalized,perdew1996phys} or PBE+$U$.

\begin{figure*}[th]
\centering 
\includegraphics[scale=0.55]{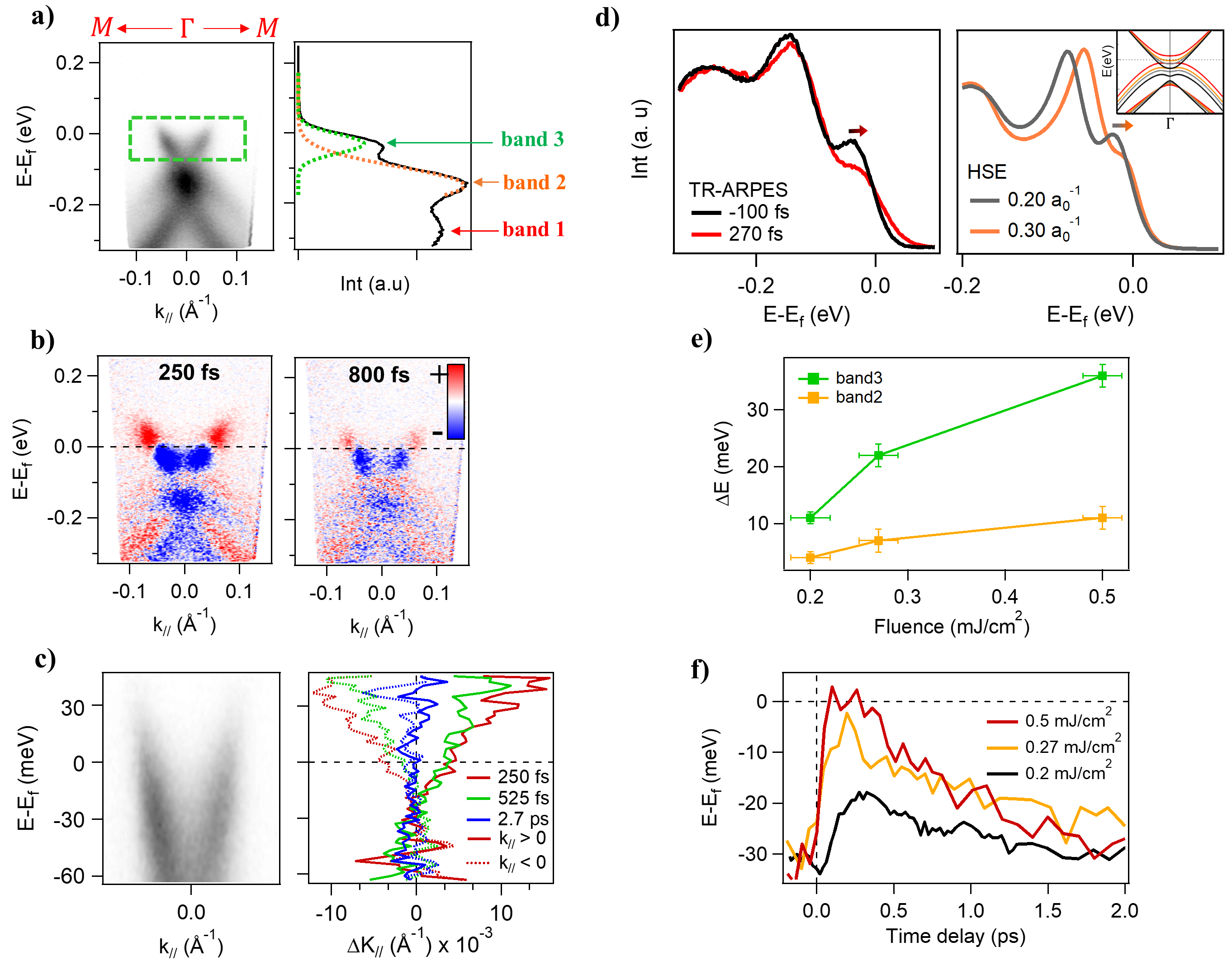}
\caption{
(a) ARPES reference image at the $\Gamma$ point obtained with \textit{s}-polarization of light. Right image shows the photoemission intensity for a region of 0.2 $\AA^{-1}$ around $\Gamma$. The contribution of each of the three bands is labeled. Band number 3 generates the electron pocket. (b) Difference tr-ARPES spectra before and after photoexcitation with pump fluence of 0.2 mJ/cm$^2$. (c) Left panel shows the electron pocket while its renormalization for different time delays is shown on the right. (d) On the left panel photoemission intensity for the negative delay and 250 fs are shown. The arrow shows the direction of the shift for the electron pocket. This shift corresponds to an increase of screening effects as is reproduced by modified HSE calculations and depicted on the right panel for two different screening lengths. (e) The maximum photoinduced shift of the electron pocket (band 3) and the adjacent band (band 2) are shown as a function of pump fluence. The error bars are estimated from the experiments and fitting procedure. The gap separating band 3 and band 2 increases with excitation density. (f) Dynamics of the center of the electron pocket for various fluences, that can reach the Fermi level for the highest fluence applied in our experiments.}
\label{fig:Figure4}
\end{figure*}

By means of time-dependent DFT (TDDFT) \cite{volkov2019attosecond}, TDDFT+$U$ \cite{tancogne2018ultrafast} and GW+DMFT\cite{golevz2019dynamics,golez2019multi} frameworks, recent theoretical studies demonstrated that the changes of electronic structure induced by a light pulse in a correlated electron system are as follows: (i) change of filling; (ii) band shifts due to variations of the Hartree potential; (iii) electronic screening enhancement. In case of strong correlations, the latter change should be the most relevant one \cite{ayral2017influence}. Indeed, photo-excitation is supposed to significantly reduce the value of the local Hubbard repulsion $U$\cite{tancogne2018ultrafast,golez2019multi}. Out-of-equilibrium GW+DMFT, one of the most extended approaches, is able to treat not only strong local Coulomb repulsions, but also non-local correlation effects. However, this approach has several technical difficulties which prevent its application to real systems beyond simple cases, e.g. few-orbitals $d$-$p$ systems~\cite{golez2019multi}. 
Here, we follow a simpler scheme by letting the photo-excitation modify the screening length of the non-local Fock operator. We assume that the screening enhancement is maximum right after the pump pulse and then it progressively decays as the system relaxes to equilibrium. This dynamics can be mapped into equilibrium \emph{ab initio} calculations, provided the screening parameters are modified, for they evolve with the pump-probe time delay. As we shall see, this "instantaneous equilibrium" approximation is justified in presence of a slow relaxation rate as observed in this Dirac state (see Fig.~\ref{fig:Figure2}(g)). The HSE functional is particularly convenient in this regard, as it explicitly includes the screened exchange in a truly non-local fashion. The PBE exchange-correlation ($xc$) functional is modified by the addition of the screened interaction, treated at the Fock level ($E_{x}^{HF,\textrm{screened}}$), such that the resulting functional reads as $E_{xc}^{HSE}=E_{xc}^{PBE}+\alpha\left(E_{x}^{HF,\textrm{screened}}(\omega)-E_{x}^{PBE,\textrm{screened}}(\omega)\right)$. In our case $\alpha=0.07$, as determined in BaNiS$_2$ by comparison with quantum oscillations \cite{Klein2018a}. The screened interaction is written as: 

\begin{equation}
V^\textrm{screened}(r)=\erfc(\omega r)/r,
\end{equation}

where $\erfc$ is the complementary error function and $\omega$ is interpreted as the inverse of a screening length, $\omega=1/\lambda$. For the system at equilibrium, $\omega=0.108$ in atomic units, \emph{i.e.} the HSE regular value. In the case of a photoexcited system, $\omega$ increases and reaches its maximum just after the pump pulse. We notice that the band structure saturates for $\omega \geq 0.6$ (Fig.~\ref{fig:theory}). Correspondingly, saturation is also reached in the pump-probe experiment for very large fluence values of 0.5 mJ/cm$^2$. For these values, the integrated density of states at the Fermi level is the largest, c.f. SM. Within the validity of the Thomas-Fermi model, this implies that the screening length is the shortest, \emph{i.e.} the screening is enhanced.

We performed \emph{ab initio} fully relativistic calculations using the modified HSE functional for $\alpha=0.07$ and $\omega=\{0.108, 0.2, 0.3, 0.6\}$ by using the \textsc{Quantum Espresso} package \cite{QE-2009,giannozzi2017}. The geometry is taken from the experiment, as described by Grey \emph{et al.} \cite{Grey1970}. Further details of the calculations are explained in the SM. 

The calculated low-energy band structure is shown in Fig.~\ref{fig:theory}(a). The full bandwidth is reduced by the screening enhancement, which contrasts the well-known tendency of the Coulomb exchange to widen the bandwidth. The impact on the band renormalization of the Dirac states is reported in Fig.~\ref{fig:theory}(b). According to our calculations, the Fermi velocities are reduced by $\approx$ 10 \% by photoexcitation, in both Dirac cone branches of $d_{z^2}$ and $d_{x^2-y^2}$ character, respectively (c.f. SM). Experimentally, the Fermi velocity reduction is even larger and reaches 30 \% (Fig.~\ref{fig:Figure2}(h)), in both sides of the cone. This discrepancy is attributed to the dynamical nature of photoexcitation and to further correlation effects beyond the screened exchange, neglected in our scheme.

In order to further verify the agreement between theory and experiment, we analyze the expected behavior of the band structure around the $\Gamma$ point upon photoexcitation. The \emph{ab initio} results are plotted in Fig.~\ref{fig:theory}(c). In the enhanced screening regime, the electron pocket at $\Gamma$ disappears because the generating band is shifted above the Fermi level. This subtle effect depends on the combination of enhanced screening and spin-orbit coupling, which splits the set of three bands just below the $\Gamma$ point in the system at equilibrium. The bottom band is the least touched by the screening modification, as it keeps almost the same position in energy. The very different dynamics of these three bands upon photoexcitation, as revealed by our calculations, is a stringent test for the applicability and reliability of our approach to describe the out-of-equilibrium band structure.  

The experimental results for the $\Gamma$ point are presented in Fig.~\ref{fig:Figure4}: they confirm that photoexcitation not only affects the velocity of the Dirac states but also the energy levels of the bands at $\Gamma$. Fig.~\ref{fig:Figure4}(a) shows the ARPES spectrum at $\Gamma$ along the $\Gamma - M$ direction taken with $s-$polarization. The difference ARPES images for the pump fluence of 0.2 mJ/cm$^2$ are shown in Fig.~\ref{fig:Figure4}(b). 
To fully capture the dynamics of the bands, here we focus on the electron pocket that crosses the Fermi level, shown with a dashed contour in Fig.~\ref{fig:Figure4}(a). In Fig.~\ref{fig:Figure4}(c), we present the change in the band dispersion of the electron pocket obtained by MDC analysis of the ARPES spectra for different time delays. Note a reduction of band velocity up to 250 fs and a non-rigid shift of the wave vector that is significantly more sizeable for the states above the Fermi level, i.e., around 30 meV it can reach up to 0.01 \AA$^{-1}$.

We now study the dynamics of the band structure by analyzing the energy distribution curves (EDC's). The photoemission intensity obtained by integrating the EDC's in a region of 0.2 \AA$^{-1}$ around the $\Gamma$ point is shown in Fig.~\ref{fig:Figure4}(a). The photoemission peak of the three observed bands are numbered as follows. Band number 3 refers to the electron pocket; bands 2 and 1 are the underlying bands at higher binding energies, respectively. As apparent in panel \ref{fig:Figure4}(d), the shift of band structure at the $\Gamma$ point matches well our modified HSE calculations that take into account the influence of screening. Hence, we now follow the photoinduced energy shift of the electron pocket and the neighboring band as a function of time. In Fig.~\ref{fig:Figure4}(e) we illustrate the maximum of the photoinduced energy shift at a delay of 250 fs, for the electron pocket and for the adjacent band to it (band 2). Note that the maximum of $\Delta E$ for the electron pocket reaches about 11 meV for a fluence of 0.2 mJ/cm$^2$ and reaches 36 meV for 0.5 mJ/cm$^2$. Our data give strong evidence of a non-rigid and fluence-dependent energy shift of the bands at the $\Gamma$ point. The results also demonstrate that not only the energy separating the bands close to the Fermi level is enhanced in the out-of-equilibrium state, but it is also possible to 
completely empty electron pockets using photoexcitation. This idea is further supported by tracking solely the energy position of the electron pocket for different pump powers in Fig.~\ref{fig:Figure4}(f).
In Fig.~\ref{fig:theory}(b) we clearly see that with the increase of screening, experimentally achieved by transiently driving the electrons out-of-equilibrium, the band velocity of the electron pocket decreases and its energy level moves gradually to lower binding energies and eventually crosses the Fermi level. The dynamics of the band structure reveals that while the electron pocket is highly influenced by the ultrafast photoexcitation, the rest of the bands remain rather stiff. For instance, band 1 hardly exhibits any appreciable energy shift (c.f. SM).

In conclusion, we observed that photo-excitation modifies significantly the electronic band structure of the quasi 2D Dirac semimetal BaNiS$_2$. Notable effects are a reduction of the band velocity of the Dirac states and of the binding energy of the bands at the center of the Brillouin zone. These effects are theoretically explained by a dynamical change of the screening length of non-local interactions. Namely, the enhancement of screening upon photo-excitation drives a non-rigid shift of the band structure that progressively decays to the equilibrium state upon electron relaxation. We envisage that the out-of-equilibrium phenomena reported in the present work should occur in other quasi 2D systems with sizable long-range correlations. 

\acknowledgments
M.C. acknowledges the Grand \'Equipement National de Calcul Intensif (GENCI) for providing the necessary computational time to carry out the calculations presented in this work, through the project number 0906493. 
This work was supported by the Region Ile-de-France (DIM OxyMORE), the EU/FP7 under the contract Go Fast (Grant No. 280555), by "Investissement d'Avenir" Labex PALM (ANR-10-LABX-0039-PALM), by the  Equipex ATTOLAB (ANR11-EQPX0005-ATTOLAB) and by the ANR Iridoti (Grant ANR-13-IS04-0001).

\bibliography{biblio}

\begin{thebibliography}{37}%
\makeatletter
\providecommand \@ifxundefined [1]{%
 \@ifx{#1\undefined}
}%
\providecommand \@ifnum [1]{%
 \ifnum #1\expandafter \@firstoftwo
 \else \expandafter \@secondoftwo
 \fi
}%
\providecommand \@ifx [1]{%
 \ifx #1\expandafter \@firstoftwo
 \else \expandafter \@secondoftwo
 \fi
}%
\providecommand \natexlab [1]{#1}%
\providecommand \enquote  [1]{``#1''}%
\providecommand \bibnamefont  [1]{#1}%
\providecommand \bibfnamefont [1]{#1}%
\providecommand \citenamefont [1]{#1}%
\providecommand \href@noop [0]{\@secondoftwo}%
\providecommand \href [0]{\begingroup \@sanitize@url \@href}%
\providecommand \@href[1]{\@@startlink{#1}\@@href}%
\providecommand \@@href[1]{\endgroup#1\@@endlink}%
\providecommand \@sanitize@url [0]{\catcode `\\12\catcode `\$12\catcode
  `\&12\catcode `\#12\catcode `\^12\catcode `\_12\catcode `\%12\relax}%
\providecommand \@@startlink[1]{}%
\providecommand \@@endlink[0]{}%
\providecommand \url  [0]{\begingroup\@sanitize@url \@url }%
\providecommand \@url [1]{\endgroup\@href {#1}{\urlprefix }}%
\providecommand \urlprefix  [0]{URL }%
\providecommand \Eprint [0]{\href }%
\providecommand \doibase [0]{http://dx.doi.org/}%
\providecommand \selectlanguage [0]{\@gobble}%
\providecommand \bibinfo  [0]{\@secondoftwo}%
\providecommand \bibfield  [0]{\@secondoftwo}%
\providecommand \translation [1]{[#1]}%
\providecommand \BibitemOpen [0]{}%
\providecommand \bibitemStop [0]{}%
\providecommand \bibitemNoStop [0]{.\EOS\space}%
\providecommand \EOS [0]{\spacefactor3000\relax}%
\providecommand \BibitemShut  [1]{\csname bibitem#1\endcsname}%
\let\auto@bib@innerbib\@empty
\bibitem [{\citenamefont {Armitage}\ \emph {et~al.}(2018)\citenamefont
  {Armitage}, \citenamefont {Mele},\ and\ \citenamefont
  {Vishwanath}}]{Armitage2018}%
  \BibitemOpen
  \bibfield  {author} {\bibinfo {author} {\bibfnamefont {N.~P.}\ \bibnamefont
  {Armitage}}, \bibinfo {author} {\bibfnamefont {E.~J.}\ \bibnamefont {Mele}},
  \ and\ \bibinfo {author} {\bibfnamefont {A.}~\bibnamefont {Vishwanath}},\
  }\href {\doibase 10.1103/RevModPhys.90.015001} {\bibfield  {journal}
  {\bibinfo  {journal} {Rev. Mod. Phys.}\ }\textbf {\bibinfo {volume} {90}},\
  \bibinfo {pages} {15001} (\bibinfo {year} {2018})}\BibitemShut {NoStop}%
\bibitem [{\citenamefont {Liu}\ \emph {et~al.}(2014)\citenamefont {Liu},
  \citenamefont {Zhou}, \citenamefont {Zhang}, \citenamefont {Wang},
  \citenamefont {Weng}, \citenamefont {Prabhakaran}, \citenamefont {Mo},
  \citenamefont {Shen}, \citenamefont {Fang}, \citenamefont {Dai},
  \citenamefont {Hussain},\ and\ \citenamefont {Chen}}]{Liu2014a}%
  \BibitemOpen
  \bibfield  {author} {\bibinfo {author} {\bibfnamefont {Z.~K.}\ \bibnamefont
  {Liu}}, \bibinfo {author} {\bibfnamefont {B.}~\bibnamefont {Zhou}}, \bibinfo
  {author} {\bibfnamefont {Y.}~\bibnamefont {Zhang}}, \bibinfo {author}
  {\bibfnamefont {Z.~J.}\ \bibnamefont {Wang}}, \bibinfo {author}
  {\bibfnamefont {H.~M.}\ \bibnamefont {Weng}}, \bibinfo {author}
  {\bibfnamefont {D.}~\bibnamefont {Prabhakaran}}, \bibinfo {author}
  {\bibfnamefont {S.-K.}\ \bibnamefont {Mo}}, \bibinfo {author} {\bibfnamefont
  {Z.~X.}\ \bibnamefont {Shen}}, \bibinfo {author} {\bibfnamefont
  {Z.}~\bibnamefont {Fang}}, \bibinfo {author} {\bibfnamefont {X.}~\bibnamefont
  {Dai}}, \bibinfo {author} {\bibfnamefont {Z.}~\bibnamefont {Hussain}}, \ and\
  \bibinfo {author} {\bibfnamefont {Y.~L.}\ \bibnamefont {Chen}},\ }\href
  {\doibase 10.1038/nmat1849} {\bibfield  {journal} {\bibinfo  {journal}
  {Science}\ }\textbf {\bibinfo {volume} {343}},\ \bibinfo {pages} {864}
  (\bibinfo {year} {2014})}\BibitemShut {NoStop}%
\bibitem [{\citenamefont {Xu}\ \emph {et~al.}(2015)\citenamefont {Xu},
  \citenamefont {Belopolski}, \citenamefont {Alidoust}, \citenamefont
  {Neupane}, \citenamefont {Bian}, \citenamefont {Zhang}, \citenamefont
  {Sankar}, \citenamefont {Chang}, \citenamefont {Yuan}, \citenamefont {Lee},
  \citenamefont {Huang}, \citenamefont {Zheng}, \citenamefont {Sanchez},
  \citenamefont {Wang}, \citenamefont {Bansil}, \citenamefont {Chou},
  \citenamefont {Shibayev}, \citenamefont {Lin}, \citenamefont {Jia},\ and\
  \citenamefont {Hasan}}]{Xu2015}%
  \BibitemOpen
  \bibfield  {author} {\bibinfo {author} {\bibfnamefont {S.-Y.}\ \bibnamefont
  {Xu}}, \bibinfo {author} {\bibfnamefont {I.}~\bibnamefont {Belopolski}},
  \bibinfo {author} {\bibfnamefont {N.}~\bibnamefont {Alidoust}}, \bibinfo
  {author} {\bibfnamefont {M.}~\bibnamefont {Neupane}}, \bibinfo {author}
  {\bibfnamefont {G.}~\bibnamefont {Bian}}, \bibinfo {author} {\bibfnamefont
  {C.}~\bibnamefont {Zhang}}, \bibinfo {author} {\bibfnamefont
  {R.}~\bibnamefont {Sankar}}, \bibinfo {author} {\bibfnamefont
  {G.}~\bibnamefont {Chang}}, \bibinfo {author} {\bibfnamefont
  {Z.}~\bibnamefont {Yuan}}, \bibinfo {author} {\bibfnamefont {C.-C.}\
  \bibnamefont {Lee}}, \bibinfo {author} {\bibfnamefont {S.-M.}\ \bibnamefont
  {Huang}}, \bibinfo {author} {\bibfnamefont {H.}~\bibnamefont {Zheng}},
  \bibinfo {author} {\bibfnamefont {D.~S.}\ \bibnamefont {Sanchez}}, \bibinfo
  {author} {\bibfnamefont {B.}~\bibnamefont {Wang}}, \bibinfo {author}
  {\bibfnamefont {A.}~\bibnamefont {Bansil}}, \bibinfo {author} {\bibfnamefont
  {F.}~\bibnamefont {Chou}}, \bibinfo {author} {\bibfnamefont {P.~P.}\
  \bibnamefont {Shibayev}}, \bibinfo {author} {\bibfnamefont {H.}~\bibnamefont
  {Lin}}, \bibinfo {author} {\bibfnamefont {S.}~\bibnamefont {Jia}}, \ and\
  \bibinfo {author} {\bibfnamefont {M.~Z.}\ \bibnamefont {Hasan}},\ }\href@noop
  {} {\bibfield  {journal} {\bibinfo  {journal} {Science}\ }\textbf {\bibinfo
  {volume} {349}},\ \bibinfo {pages} {613} (\bibinfo {year}
  {2015})}\BibitemShut {NoStop}%
\bibitem [{\citenamefont {Wang}\ \emph {et~al.}(2012)\citenamefont {Wang},
  \citenamefont {Sun}, \citenamefont {Chen}, \citenamefont {Franchini},
  \citenamefont {Xu}, \citenamefont {Weng}, \citenamefont {Dai},\ and\
  \citenamefont {Fang}}]{Wang2012}%
  \BibitemOpen
  \bibfield  {author} {\bibinfo {author} {\bibfnamefont {Z.}~\bibnamefont
  {Wang}}, \bibinfo {author} {\bibfnamefont {Y.}~\bibnamefont {Sun}}, \bibinfo
  {author} {\bibfnamefont {X.~Q.}\ \bibnamefont {Chen}}, \bibinfo {author}
  {\bibfnamefont {C.}~\bibnamefont {Franchini}}, \bibinfo {author}
  {\bibfnamefont {G.}~\bibnamefont {Xu}}, \bibinfo {author} {\bibfnamefont
  {H.}~\bibnamefont {Weng}}, \bibinfo {author} {\bibfnamefont {X.}~\bibnamefont
  {Dai}}, \ and\ \bibinfo {author} {\bibfnamefont {Z.}~\bibnamefont {Fang}},\
  }\href {\doibase 10.1103/PhysRevB.85.195320} {\bibfield  {journal} {\bibinfo
  {journal} {Phys. Rev. B}\ }\textbf {\bibinfo {volume} {85}},\ \bibinfo
  {pages} {195320} (\bibinfo {year} {2012})}\BibitemShut {NoStop}%
\bibitem [{\citenamefont {Borisenko}\ \emph {et~al.}(2014)\citenamefont
  {Borisenko}, \citenamefont {Gibson}, \citenamefont {Evtushinsky},
  \citenamefont {Zabolotnyy}, \citenamefont {B{\"u}chner},\ and\ \citenamefont
  {Cava}}]{Borisenko2014}%
  \BibitemOpen
  \bibfield  {author} {\bibinfo {author} {\bibfnamefont {S.}~\bibnamefont
  {Borisenko}}, \bibinfo {author} {\bibfnamefont {Q.}~\bibnamefont {Gibson}},
  \bibinfo {author} {\bibfnamefont {D.}~\bibnamefont {Evtushinsky}}, \bibinfo
  {author} {\bibfnamefont {V.}~\bibnamefont {Zabolotnyy}}, \bibinfo {author}
  {\bibfnamefont {B.}~\bibnamefont {B{\"u}chner}}, \ and\ \bibinfo {author}
  {\bibfnamefont {R.~J.}\ \bibnamefont {Cava}},\ }\href@noop {} {\bibfield
  {journal} {\bibinfo  {journal} {Phys. Rev. Lett.}\ }\textbf {\bibinfo
  {volume} {113}},\ \bibinfo {pages} {027603} (\bibinfo {year}
  {2014})}\BibitemShut {NoStop}%
\bibitem [{\citenamefont {Wang}\ \emph {et~al.}(2013)\citenamefont {Wang},
  \citenamefont {Steinberg}, \citenamefont {Jarillo-Herrero},\ and\
  \citenamefont {Gedik}}]{Floquet2013}%
  \BibitemOpen
  \bibfield  {author} {\bibinfo {author} {\bibfnamefont {Y.~H.}\ \bibnamefont
  {Wang}}, \bibinfo {author} {\bibfnamefont {H.}~\bibnamefont {Steinberg}},
  \bibinfo {author} {\bibfnamefont {P.}~\bibnamefont {Jarillo-Herrero}}, \ and\
  \bibinfo {author} {\bibfnamefont {N.}~\bibnamefont {Gedik}},\ }\href
  {\doibase 10.1126/science.1239834} {\bibfield  {journal} {\bibinfo  {journal}
  {Science}\ }\textbf {\bibinfo {volume} {342}},\ \bibinfo {pages} {453}
  (\bibinfo {year} {2013})}\BibitemShut {NoStop}%
\bibitem [{\citenamefont {Weber}\ \emph {et~al.}(2017)\citenamefont {Weber},
  \citenamefont {Berggren}, \citenamefont {Masten}, \citenamefont {Ogloza},
  \citenamefont {Deckoff-Jones}, \citenamefont {Mad{\'{e}}o}, \citenamefont
  {Man}, \citenamefont {Dani}, \citenamefont {Zhao}, \citenamefont {Chen},
  \citenamefont {Liu}, \citenamefont {Mao}, \citenamefont {Schoop},
  \citenamefont {Lotsch}, \citenamefont {Parkin},\ and\ \citenamefont
  {Mazhar}}]{Weber2017}%
  \BibitemOpen
  \bibfield  {author} {\bibinfo {author} {\bibfnamefont {C.~P.}\ \bibnamefont
  {Weber}}, \bibinfo {author} {\bibfnamefont {B.~S.}\ \bibnamefont {Berggren}},
  \bibinfo {author} {\bibfnamefont {M.~G.}\ \bibnamefont {Masten}}, \bibinfo
  {author} {\bibfnamefont {T.~C.}\ \bibnamefont {Ogloza}}, \bibinfo {author}
  {\bibfnamefont {S.}~\bibnamefont {Deckoff-Jones}}, \bibinfo {author}
  {\bibfnamefont {J.}~\bibnamefont {Mad{\'{e}}o}}, \bibinfo {author}
  {\bibfnamefont {M.~K.~L.}\ \bibnamefont {Man}}, \bibinfo {author}
  {\bibfnamefont {K.~M.}\ \bibnamefont {Dani}}, \bibinfo {author}
  {\bibfnamefont {L.}~\bibnamefont {Zhao}}, \bibinfo {author} {\bibfnamefont
  {G.}~\bibnamefont {Chen}}, \bibinfo {author} {\bibfnamefont {J.}~\bibnamefont
  {Liu}}, \bibinfo {author} {\bibfnamefont {Z.}~\bibnamefont {Mao}}, \bibinfo
  {author} {\bibfnamefont {L.~M.}\ \bibnamefont {Schoop}}, \bibinfo {author}
  {\bibfnamefont {B.~V.}\ \bibnamefont {Lotsch}}, \bibinfo {author}
  {\bibfnamefont {S.~S.~P.}\ \bibnamefont {Parkin}}, \ and\ \bibinfo {author}
  {\bibfnamefont {A.}~\bibnamefont {Mazhar}},\ }\href@noop {} {\bibfield
  {journal} {\bibinfo  {journal} {J. Appl. Phys.}\ }\textbf {\bibinfo {volume}
  {122}},\ \bibinfo {pages} {223102} (\bibinfo {year} {2017})}\BibitemShut
  {NoStop}%
\bibitem [{\citenamefont {Sie}\ \emph {et~al.}(2019)\citenamefont {Sie},
  \citenamefont {Nyby}, \citenamefont {Pemmaraju}, \citenamefont {Park},
  \citenamefont {Shen}, \citenamefont {Yang}, \citenamefont {Hoffmann},
  \citenamefont {Ofori-Okai}, \citenamefont {Li}, \citenamefont {Reid},
  \citenamefont {Weathersby}, \citenamefont {Mannebach}, \citenamefont
  {Finney}, \citenamefont {Rhodes}, \citenamefont {Chenet}, \citenamefont
  {Antony}, \citenamefont {Balicas}, \citenamefont {Hone}, \citenamefont
  {Devereaux}, \citenamefont {Heinz}, \citenamefont {Wang},\ and\ \citenamefont
  {Lindenberg}}]{Sie2019}%
  \BibitemOpen
  \bibfield  {author} {\bibinfo {author} {\bibfnamefont {E.~J.}\ \bibnamefont
  {Sie}}, \bibinfo {author} {\bibfnamefont {C.~M.}\ \bibnamefont {Nyby}},
  \bibinfo {author} {\bibfnamefont {C.~D.}\ \bibnamefont {Pemmaraju}}, \bibinfo
  {author} {\bibfnamefont {S.~J.}\ \bibnamefont {Park}}, \bibinfo {author}
  {\bibfnamefont {X.}~\bibnamefont {Shen}}, \bibinfo {author} {\bibfnamefont
  {J.}~\bibnamefont {Yang}}, \bibinfo {author} {\bibfnamefont {M.~C.}\
  \bibnamefont {Hoffmann}}, \bibinfo {author} {\bibfnamefont {B.~K.}\
  \bibnamefont {Ofori-Okai}}, \bibinfo {author} {\bibfnamefont
  {R.}~\bibnamefont {Li}}, \bibinfo {author} {\bibfnamefont {A.~H.}\
  \bibnamefont {Reid}}, \bibinfo {author} {\bibfnamefont {S.}~\bibnamefont
  {Weathersby}}, \bibinfo {author} {\bibfnamefont {E.}~\bibnamefont
  {Mannebach}}, \bibinfo {author} {\bibfnamefont {N.}~\bibnamefont {Finney}},
  \bibinfo {author} {\bibfnamefont {D.}~\bibnamefont {Rhodes}}, \bibinfo
  {author} {\bibfnamefont {D.}~\bibnamefont {Chenet}}, \bibinfo {author}
  {\bibfnamefont {A.}~\bibnamefont {Antony}}, \bibinfo {author} {\bibfnamefont
  {L.}~\bibnamefont {Balicas}}, \bibinfo {author} {\bibfnamefont
  {J.}~\bibnamefont {Hone}}, \bibinfo {author} {\bibfnamefont {T.~P.}\
  \bibnamefont {Devereaux}}, \bibinfo {author} {\bibfnamefont {T.~F.}\
  \bibnamefont {Heinz}}, \bibinfo {author} {\bibfnamefont {X.}~\bibnamefont
  {Wang}}, \ and\ \bibinfo {author} {\bibfnamefont {A.~M.}\ \bibnamefont
  {Lindenberg}},\ }\href {\doibase 10.1038/s41586-018-0809-4} {\bibfield
  {journal} {\bibinfo  {journal} {Nature}\ }\textbf {\bibinfo {volume} {565}},\
  \bibinfo {pages} {61} (\bibinfo {year} {2019})}\BibitemShut {NoStop}%
\bibitem [{\citenamefont {Young}\ \emph {et~al.}(2012)\citenamefont {Young},
  \citenamefont {Zaheer}, \citenamefont {Teo}, \citenamefont {Kane},
  \citenamefont {Mele},\ and\ \citenamefont {Rappe}}]{Young2012}%
  \BibitemOpen
  \bibfield  {author} {\bibinfo {author} {\bibfnamefont {S.~M.}\ \bibnamefont
  {Young}}, \bibinfo {author} {\bibfnamefont {S.}~\bibnamefont {Zaheer}},
  \bibinfo {author} {\bibfnamefont {J.~C.~Y.}\ \bibnamefont {Teo}}, \bibinfo
  {author} {\bibfnamefont {C.~L.}\ \bibnamefont {Kane}}, \bibinfo {author}
  {\bibfnamefont {E.~J.}\ \bibnamefont {Mele}}, \ and\ \bibinfo {author}
  {\bibfnamefont {A.~M.}\ \bibnamefont {Rappe}},\ }\href {\doibase
  10.1103/PhysRevLett.108.140405} {\bibfield  {journal} {\bibinfo  {journal}
  {Phys. Rev. Lett.}\ }\textbf {\bibinfo {volume} {108}},\ \bibinfo {pages}
  {140405} (\bibinfo {year} {2012})}\BibitemShut {NoStop}%
\bibitem [{\citenamefont {Young}\ and\ \citenamefont {Kane}(2015)}]{Young2015}%
  \BibitemOpen
  \bibfield  {author} {\bibinfo {author} {\bibfnamefont {S.~M.}\ \bibnamefont
  {Young}}\ and\ \bibinfo {author} {\bibfnamefont {C.~L.}\ \bibnamefont
  {Kane}},\ }\href {\doibase 10.1103/PhysRevLett.115.126803} {\bibfield
  {journal} {\bibinfo  {journal} {Phys. Rev. Lett.}\ }\textbf {\bibinfo
  {volume} {155}},\ \bibinfo {pages} {126803} (\bibinfo {year}
  {2015})}\BibitemShut {NoStop}%
\bibitem [{\citenamefont {Tarruell}\ \emph {et~al.}(2012)\citenamefont
  {Tarruell}, \citenamefont {Greif}, \citenamefont {Uehlinger}, \citenamefont
  {Jotzu},\ and\ \citenamefont {Esslinger}}]{Tar12}%
  \BibitemOpen
  \bibfield  {author} {\bibinfo {author} {\bibfnamefont {L.}~\bibnamefont
  {Tarruell}}, \bibinfo {author} {\bibfnamefont {D.}~\bibnamefont {Greif}},
  \bibinfo {author} {\bibfnamefont {T.}~\bibnamefont {Uehlinger}}, \bibinfo
  {author} {\bibfnamefont {G.}~\bibnamefont {Jotzu}}, \ and\ \bibinfo {author}
  {\bibfnamefont {T.}~\bibnamefont {Esslinger}},\ }\href@noop {} {\bibfield
  {journal} {\bibinfo  {journal} {Nature}\ }\textbf {\bibinfo {volume} {483}},\
  \bibinfo {pages} {302} (\bibinfo {year} {2012})}\BibitemShut {NoStop}%
\bibitem [{\citenamefont {Bahramy}\ \emph {et~al.}(2017)\citenamefont
  {Bahramy}, \citenamefont {Clark}, \citenamefont {Yang}, \citenamefont {Feng},
  \citenamefont {Bawden}, \citenamefont {Riley}, \citenamefont {Markovic},
  \citenamefont {Mazzola}, \citenamefont {Sunko}, \citenamefont {Biswas},
  \citenamefont {Cooil}, \citenamefont {Jorge}, \citenamefont {Wells},
  \citenamefont {Leandersson}, \citenamefont {Balasubramanian}, \citenamefont
  {Fujii}, \citenamefont {Vobornik}, \citenamefont {Rault}, \citenamefont
  {Kim}, \citenamefont {Hoesch}, \citenamefont {Okawa}, \citenamefont
  {Asakawa}, \citenamefont {Sasagawa}, \citenamefont {Eknapakul}, \citenamefont
  {Meevasana},\ and\ \citenamefont {King}}]{Bahramy2017}%
  \BibitemOpen
  \bibfield  {author} {\bibinfo {author} {\bibfnamefont {M.~S.}\ \bibnamefont
  {Bahramy}}, \bibinfo {author} {\bibfnamefont {O.~J.}\ \bibnamefont {Clark}},
  \bibinfo {author} {\bibfnamefont {B.~J.}\ \bibnamefont {Yang}}, \bibinfo
  {author} {\bibfnamefont {J.}~\bibnamefont {Feng}}, \bibinfo {author}
  {\bibfnamefont {L.}~\bibnamefont {Bawden}}, \bibinfo {author} {\bibfnamefont
  {J.~M.}\ \bibnamefont {Riley}}, \bibinfo {author} {\bibfnamefont
  {I.}~\bibnamefont {Markovic}}, \bibinfo {author} {\bibfnamefont
  {F.}~\bibnamefont {Mazzola}}, \bibinfo {author} {\bibfnamefont
  {V.}~\bibnamefont {Sunko}}, \bibinfo {author} {\bibfnamefont
  {D.}~\bibnamefont {Biswas}}, \bibinfo {author} {\bibfnamefont {S.~P.}\
  \bibnamefont {Cooil}}, \bibinfo {author} {\bibfnamefont {M.}~\bibnamefont
  {Jorge}}, \bibinfo {author} {\bibfnamefont {J.~W.}\ \bibnamefont {Wells}},
  \bibinfo {author} {\bibfnamefont {M.}~\bibnamefont {Leandersson}}, \bibinfo
  {author} {\bibfnamefont {T.}~\bibnamefont {Balasubramanian}}, \bibinfo
  {author} {\bibfnamefont {J.}~\bibnamefont {Fujii}}, \bibinfo {author}
  {\bibfnamefont {I.}~\bibnamefont {Vobornik}}, \bibinfo {author}
  {\bibfnamefont {J.~E.}\ \bibnamefont {Rault}}, \bibinfo {author}
  {\bibfnamefont {T.~K.}\ \bibnamefont {Kim}}, \bibinfo {author} {\bibfnamefont
  {M.}~\bibnamefont {Hoesch}}, \bibinfo {author} {\bibfnamefont
  {K.}~\bibnamefont {Okawa}}, \bibinfo {author} {\bibfnamefont
  {M.}~\bibnamefont {Asakawa}}, \bibinfo {author} {\bibfnamefont
  {T.}~\bibnamefont {Sasagawa}}, \bibinfo {author} {\bibfnamefont
  {T.}~\bibnamefont {Eknapakul}}, \bibinfo {author} {\bibfnamefont
  {W.}~\bibnamefont {Meevasana}}, \ and\ \bibinfo {author} {\bibfnamefont
  {P.~D.}\ \bibnamefont {King}},\ }\href {\doibase 10.1038/NMAT5031} {\bibfield
   {journal} {\bibinfo  {journal} {Nat. Mater.}\ }\textbf {\bibinfo {volume}
  {17}},\ \bibinfo {pages} {21} (\bibinfo {year} {2017})}\BibitemShut {NoStop}%
\bibitem [{\citenamefont {Santos-Cottin}\ \emph {et~al.}(2016)\citenamefont
  {Santos-Cottin}, \citenamefont {Casula}, \citenamefont {Lantz}, \citenamefont
  {Klein}, \citenamefont {Petaccia}, \citenamefont {Le~F\`{e}vre},
  \citenamefont {Bertran}, \citenamefont {Papalazarou}, \citenamefont {Marsi},\
  and\ \citenamefont {Gauzzi}}]{SantosCottin2016}%
  \BibitemOpen
  \bibfield  {author} {\bibinfo {author} {\bibfnamefont {D.}~\bibnamefont
  {Santos-Cottin}}, \bibinfo {author} {\bibfnamefont {M.}~\bibnamefont
  {Casula}}, \bibinfo {author} {\bibfnamefont {G.}~\bibnamefont {Lantz}},
  \bibinfo {author} {\bibfnamefont {Y.}~\bibnamefont {Klein}}, \bibinfo
  {author} {\bibfnamefont {L.}~\bibnamefont {Petaccia}}, \bibinfo {author}
  {\bibfnamefont {P.}~\bibnamefont {Le~F\`{e}vre}}, \bibinfo {author}
  {\bibfnamefont {F.}~\bibnamefont {Bertran}}, \bibinfo {author} {\bibfnamefont
  {E.}~\bibnamefont {Papalazarou}}, \bibinfo {author} {\bibfnamefont
  {M.}~\bibnamefont {Marsi}}, \ and\ \bibinfo {author} {\bibfnamefont
  {A.}~\bibnamefont {Gauzzi}},\ }\href@noop {} {\bibfield  {journal} {\bibinfo
  {journal} {Nat. Commun.}\ }\textbf {\bibinfo {volume} {7}},\ \bibinfo {pages}
  {11258} (\bibinfo {year} {2016})}\BibitemShut {NoStop}%
\bibitem [{\citenamefont {Brouet}\ \emph {et~al.}(2013)\citenamefont {Brouet},
  \citenamefont {Mauchain}, \citenamefont {Papalazarou}, \citenamefont {Faure},
  \citenamefont {Marsi}, \citenamefont {Lin}, \citenamefont {Taleb-Ibrahimi},
  \citenamefont {{Le F{\`{e}}vre}}, \citenamefont {Bertran}, \citenamefont
  {Cario}, \citenamefont {Janod}, \citenamefont {Corraze}, \citenamefont
  {Phuoc},\ and\ \citenamefont {Perfetti}}]{Brouet2013}%
  \BibitemOpen
  \bibfield  {author} {\bibinfo {author} {\bibfnamefont {V.}~\bibnamefont
  {Brouet}}, \bibinfo {author} {\bibfnamefont {J.}~\bibnamefont {Mauchain}},
  \bibinfo {author} {\bibfnamefont {E.}~\bibnamefont {Papalazarou}}, \bibinfo
  {author} {\bibfnamefont {J.}~\bibnamefont {Faure}}, \bibinfo {author}
  {\bibfnamefont {M.}~\bibnamefont {Marsi}}, \bibinfo {author} {\bibfnamefont
  {P.~H.}\ \bibnamefont {Lin}}, \bibinfo {author} {\bibfnamefont
  {A.}~\bibnamefont {Taleb-Ibrahimi}}, \bibinfo {author} {\bibfnamefont
  {P.}~\bibnamefont {{Le F{\`{e}}vre}}}, \bibinfo {author} {\bibfnamefont
  {F.}~\bibnamefont {Bertran}}, \bibinfo {author} {\bibfnamefont
  {L.}~\bibnamefont {Cario}}, \bibinfo {author} {\bibfnamefont
  {E.}~\bibnamefont {Janod}}, \bibinfo {author} {\bibfnamefont
  {B.}~\bibnamefont {Corraze}}, \bibinfo {author} {\bibfnamefont {V.~T.}\
  \bibnamefont {Phuoc}}, \ and\ \bibinfo {author} {\bibfnamefont
  {L.}~\bibnamefont {Perfetti}},\ }\href {\doibase 10.1103/PhysRevB.87.041106}
  {\bibfield  {journal} {\bibinfo  {journal} {Phys. Rev. B}\ }\textbf {\bibinfo
  {volume} {87}},\ \bibinfo {pages} {041106(R)} (\bibinfo {year}
  {2013})}\BibitemShut {NoStop}%
\bibitem [{\citenamefont {Rameau}\ \emph {et~al.}(2014)\citenamefont {Rameau},
  \citenamefont {Freutel}, \citenamefont {Rettig}, \citenamefont {Avigo},
  \citenamefont {Ligges}, \citenamefont {Yoshida}, \citenamefont {Eisaki},
  \citenamefont {Schneeloch}, \citenamefont {Zhong}, \citenamefont {Xu},
  \citenamefont {Gu}, \citenamefont {Johnson},\ and\ \citenamefont
  {Bovensiepen}}]{Rameau2014}%
  \BibitemOpen
  \bibfield  {author} {\bibinfo {author} {\bibfnamefont {J.~D.}\ \bibnamefont
  {Rameau}}, \bibinfo {author} {\bibfnamefont {S.}~\bibnamefont {Freutel}},
  \bibinfo {author} {\bibfnamefont {L.}~\bibnamefont {Rettig}}, \bibinfo
  {author} {\bibfnamefont {I.}~\bibnamefont {Avigo}}, \bibinfo {author}
  {\bibfnamefont {M.}~\bibnamefont {Ligges}}, \bibinfo {author} {\bibfnamefont
  {Y.}~\bibnamefont {Yoshida}}, \bibinfo {author} {\bibfnamefont
  {H.}~\bibnamefont {Eisaki}}, \bibinfo {author} {\bibfnamefont
  {J.}~\bibnamefont {Schneeloch}}, \bibinfo {author} {\bibfnamefont {R.~D.}\
  \bibnamefont {Zhong}}, \bibinfo {author} {\bibfnamefont {Z.~J.}\ \bibnamefont
  {Xu}}, \bibinfo {author} {\bibfnamefont {G.~D.}\ \bibnamefont {Gu}}, \bibinfo
  {author} {\bibfnamefont {P.~D.}\ \bibnamefont {Johnson}}, \ and\ \bibinfo
  {author} {\bibfnamefont {U.}~\bibnamefont {Bovensiepen}},\ }\href {\doibase
  10.1103/PhysRevB.89.115115} {\bibfield  {journal} {\bibinfo  {journal} {Phys.
  Rev. B}\ }\textbf {\bibinfo {volume} {89}},\ \bibinfo {pages} {115115}
  (\bibinfo {year} {2014})}\BibitemShut {NoStop}%
\bibitem [{\citenamefont {Mor}\ \emph {et~al.}(2017)\citenamefont {Mor},
  \citenamefont {Herzog}, \citenamefont {Gole}, \citenamefont {Werner},
  \citenamefont {Eckstein}, \citenamefont {Katayama}, \citenamefont {Nohara},
  \citenamefont {Takagi}, \citenamefont {Mizokawa}, \citenamefont {Monney},\
  and\ \citenamefont {St{\"{a}}hler}}]{Mor2017}%
  \BibitemOpen
  \bibfield  {author} {\bibinfo {author} {\bibfnamefont {S.}~\bibnamefont
  {Mor}}, \bibinfo {author} {\bibfnamefont {M.}~\bibnamefont {Herzog}},
  \bibinfo {author} {\bibfnamefont {D.}~\bibnamefont {Gole}}, \bibinfo {author}
  {\bibfnamefont {P.}~\bibnamefont {Werner}}, \bibinfo {author} {\bibfnamefont
  {M.}~\bibnamefont {Eckstein}}, \bibinfo {author} {\bibfnamefont
  {N.}~\bibnamefont {Katayama}}, \bibinfo {author} {\bibfnamefont
  {M.}~\bibnamefont {Nohara}}, \bibinfo {author} {\bibfnamefont
  {H.}~\bibnamefont {Takagi}}, \bibinfo {author} {\bibfnamefont
  {T.}~\bibnamefont {Mizokawa}}, \bibinfo {author} {\bibfnamefont
  {C.}~\bibnamefont {Monney}}, \ and\ \bibinfo {author} {\bibfnamefont
  {J.}~\bibnamefont {St{\"{a}}hler}},\ }\href {\doibase
  10.1103/PhysRevLett.119.086401} {\bibfield  {journal} {\bibinfo  {journal}
  {Phys. Rev. Lett.}\ }\textbf {\bibinfo {volume} {119}},\ \bibinfo {pages}
  {086401} (\bibinfo {year} {2017})}\BibitemShut {NoStop}%
\bibitem [{\citenamefont {Roth}\ \emph {et~al.}(2019)\citenamefont {Roth},
  \citenamefont {Crepaldi}, \citenamefont {Puppin}, \citenamefont {Gatti},
  \citenamefont {Bugini}, \citenamefont {Grimaldi}, \citenamefont {Barrilot},
  \citenamefont {Arrell}, \citenamefont {Frassetto}, \citenamefont {Poletto},
  \citenamefont {Chergui}, \citenamefont {Marini},\ and\ \citenamefont
  {Grioni}}]{Roth2019}%
  \BibitemOpen
  \bibfield  {author} {\bibinfo {author} {\bibfnamefont {S.}~\bibnamefont
  {Roth}}, \bibinfo {author} {\bibfnamefont {A.}~\bibnamefont {Crepaldi}},
  \bibinfo {author} {\bibfnamefont {M.}~\bibnamefont {Puppin}}, \bibinfo
  {author} {\bibfnamefont {G.}~\bibnamefont {Gatti}}, \bibinfo {author}
  {\bibfnamefont {D.}~\bibnamefont {Bugini}}, \bibinfo {author} {\bibfnamefont
  {I.}~\bibnamefont {Grimaldi}}, \bibinfo {author} {\bibfnamefont {T.~R.}\
  \bibnamefont {Barrilot}}, \bibinfo {author} {\bibfnamefont {C.~A.}\
  \bibnamefont {Arrell}}, \bibinfo {author} {\bibfnamefont {F.}~\bibnamefont
  {Frassetto}}, \bibinfo {author} {\bibfnamefont {L.}~\bibnamefont {Poletto}},
  \bibinfo {author} {\bibfnamefont {M.}~\bibnamefont {Chergui}}, \bibinfo
  {author} {\bibfnamefont {A.}~\bibnamefont {Marini}}, \ and\ \bibinfo {author}
  {\bibfnamefont {M.}~\bibnamefont {Grioni}},\ }\href@noop {} {\bibfield
  {journal} {\bibinfo  {journal} {2D Mater.}\ }\textbf {\bibinfo {volume}
  {6}},\ \bibinfo {pages} {031001} (\bibinfo {year} {2019})}\BibitemShut
  {NoStop}%
\bibitem [{\citenamefont {Ulstrup}\ \emph {et~al.}(2016)\citenamefont
  {Ulstrup}, \citenamefont {Cabo}, \citenamefont {Miwa}, \citenamefont {Riley},
  \citenamefont {Gr{\o}nborg}, \citenamefont {Johannsen}, \citenamefont
  {Cacho}, \citenamefont {Alexander}, \citenamefont {Chapman}, \citenamefont
  {Springate}, \citenamefont {Bianchi}, \citenamefont {Dendzik}, \citenamefont
  {Lauritsen}, \citenamefont {King},\ and\ \citenamefont
  {Hofmann}}]{Ulstrup2016}%
  \BibitemOpen
  \bibfield  {author} {\bibinfo {author} {\bibfnamefont {S.}~\bibnamefont
  {Ulstrup}}, \bibinfo {author} {\bibfnamefont {A.~G.}\ \bibnamefont {Cabo}},
  \bibinfo {author} {\bibfnamefont {J.~A.}\ \bibnamefont {Miwa}}, \bibinfo
  {author} {\bibfnamefont {J.~M.}\ \bibnamefont {Riley}}, \bibinfo {author}
  {\bibfnamefont {S.~S.}\ \bibnamefont {Gr{\o}nborg}}, \bibinfo {author}
  {\bibfnamefont {J.~C.}\ \bibnamefont {Johannsen}}, \bibinfo {author}
  {\bibfnamefont {C.}~\bibnamefont {Cacho}}, \bibinfo {author} {\bibfnamefont
  {O.}~\bibnamefont {Alexander}}, \bibinfo {author} {\bibfnamefont {R.~T.}\
  \bibnamefont {Chapman}}, \bibinfo {author} {\bibfnamefont {E.}~\bibnamefont
  {Springate}}, \bibinfo {author} {\bibfnamefont {M.}~\bibnamefont {Bianchi}},
  \bibinfo {author} {\bibfnamefont {M.}~\bibnamefont {Dendzik}}, \bibinfo
  {author} {\bibfnamefont {J.~V.}\ \bibnamefont {Lauritsen}}, \bibinfo {author}
  {\bibfnamefont {P.~D.~C.}\ \bibnamefont {King}}, \ and\ \bibinfo {author}
  {\bibfnamefont {P.}~\bibnamefont {Hofmann}},\ }\href {\doibase
  10.1021/acsnano.6b02622} {\bibfield  {journal} {\bibinfo  {journal} {ACS
  Nano}\ }\textbf {\bibinfo {volume} {10}},\ \bibinfo {pages} {6315} (\bibinfo
  {year} {2016})}\BibitemShut {NoStop}%
\bibitem [{\citenamefont {Nicholson}\ and\ \citenamefont
  {Wolf}(2018)}]{Nicholson2018}%
  \BibitemOpen
  \bibfield  {author} {\bibinfo {author} {\bibfnamefont {W.~G. P. M. R. L.
  E.~R.}\ \bibnamefont {Nicholson}, \bibfnamefont {C.~W.~Schmidt}}\ and\
  \bibinfo {author} {\bibfnamefont {M.}~\bibnamefont {Wolf}},\ }\href@noop {}
  {\bibfield  {journal} {\bibinfo  {journal} {Science}\ }\textbf {\bibinfo
  {volume} {362}},\ \bibinfo {pages} {821} (\bibinfo {year}
  {2018})}\BibitemShut {NoStop}%
\bibitem [{\citenamefont {Faure}\ \emph {et~al.}(2012)\citenamefont {Faure},
  \citenamefont {Mauchain}, \citenamefont {Papalazarou}, \citenamefont {Yan},
  \citenamefont {Pinon}, \citenamefont {Marsi},\ and\ \citenamefont
  {Perfetti}}]{Faure2012}%
  \BibitemOpen
  \bibfield  {author} {\bibinfo {author} {\bibfnamefont {J.}~\bibnamefont
  {Faure}}, \bibinfo {author} {\bibfnamefont {J.}~\bibnamefont {Mauchain}},
  \bibinfo {author} {\bibfnamefont {E.}~\bibnamefont {Papalazarou}}, \bibinfo
  {author} {\bibfnamefont {W.}~\bibnamefont {Yan}}, \bibinfo {author}
  {\bibfnamefont {J.}~\bibnamefont {Pinon}}, \bibinfo {author} {\bibfnamefont
  {M.}~\bibnamefont {Marsi}}, \ and\ \bibinfo {author} {\bibfnamefont
  {L.}~\bibnamefont {Perfetti}},\ }\href {\doibase 10.1063/1.3700190}
  {\bibfield  {journal} {\bibinfo  {journal} {Rev. Sci. Instrum.}\ }\textbf
  {\bibinfo {volume} {83}},\ \bibinfo {pages} {043109} (\bibinfo {year}
  {2012})}\BibitemShut {NoStop}%
\bibitem [{\citenamefont {Nilforoushan}\ \emph {et~al.}(2019)\citenamefont
  {Nilforoushan}, \citenamefont {Casula}, \citenamefont {Amaricci},
  \citenamefont {Caputo}, \citenamefont {Caillaux}, \citenamefont {Khalil},
  \citenamefont {Papalazarou}, \citenamefont {Simon}, \citenamefont {Perfetti},
  \citenamefont {Vobornik}, \citenamefont {Das}, \citenamefont {Fujii},
  \citenamefont {Santos-Cottin}, \citenamefont {Klein}, \citenamefont
  {Fabrizio}, \citenamefont {Gauzzi},\ and\ \citenamefont {Marsi}}]{arxivBNS}%
  \BibitemOpen
  \bibfield  {author} {\bibinfo {author} {\bibfnamefont {N.}~\bibnamefont
  {Nilforoushan}}, \bibinfo {author} {\bibfnamefont {M.}~\bibnamefont
  {Casula}}, \bibinfo {author} {\bibfnamefont {A.}~\bibnamefont {Amaricci}},
  \bibinfo {author} {\bibfnamefont {M.}~\bibnamefont {Caputo}}, \bibinfo
  {author} {\bibfnamefont {J.}~\bibnamefont {Caillaux}}, \bibinfo {author}
  {\bibfnamefont {L.}~\bibnamefont {Khalil}}, \bibinfo {author} {\bibfnamefont
  {E.}~\bibnamefont {Papalazarou}}, \bibinfo {author} {\bibfnamefont
  {P.}~\bibnamefont {Simon}}, \bibinfo {author} {\bibfnamefont
  {L.}~\bibnamefont {Perfetti}}, \bibinfo {author} {\bibfnamefont
  {I.}~\bibnamefont {Vobornik}}, \bibinfo {author} {\bibfnamefont {P.~K.}\
  \bibnamefont {Das}}, \bibinfo {author} {\bibfnamefont {J.}~\bibnamefont
  {Fujii}}, \bibinfo {author} {\bibfnamefont {D.}~\bibnamefont
  {Santos-Cottin}}, \bibinfo {author} {\bibfnamefont {Y.}~\bibnamefont
  {Klein}}, \bibinfo {author} {\bibfnamefont {M.}~\bibnamefont {Fabrizio}},
  \bibinfo {author} {\bibfnamefont {A.}~\bibnamefont {Gauzzi}}, \ and\ \bibinfo
  {author} {\bibfnamefont {M.}~\bibnamefont {Marsi}},\ }\href@noop {}
  {\bibfield  {journal} {\bibinfo  {journal} {arXiv Preprint arxiv:1905.12210}\
  } (\bibinfo {year} {2019})}\BibitemShut {NoStop}%
\bibitem [{\citenamefont {Caputo}\ \emph {et~al.}(2018)\citenamefont {Caputo},
  \citenamefont {Khalil}, \citenamefont {Papalazarou}, \citenamefont
  {Nilforoushan}, \citenamefont {Perfetti}, \citenamefont {Gibson},
  \citenamefont {Cava},\ and\ \citenamefont {Marsi}}]{Caputo2018}%
  \BibitemOpen
  \bibfield  {author} {\bibinfo {author} {\bibfnamefont {M.}~\bibnamefont
  {Caputo}}, \bibinfo {author} {\bibfnamefont {L.}~\bibnamefont {Khalil}},
  \bibinfo {author} {\bibfnamefont {E.}~\bibnamefont {Papalazarou}}, \bibinfo
  {author} {\bibfnamefont {N.}~\bibnamefont {Nilforoushan}}, \bibinfo {author}
  {\bibfnamefont {L.}~\bibnamefont {Perfetti}}, \bibinfo {author}
  {\bibfnamefont {Q.~D.}\ \bibnamefont {Gibson}}, \bibinfo {author}
  {\bibfnamefont {R.~J.}\ \bibnamefont {Cava}}, \ and\ \bibinfo {author}
  {\bibfnamefont {M.}~\bibnamefont {Marsi}},\ }\href {\doibase
  10.1103/PhysRevB.97.115115} {\bibfield  {journal} {\bibinfo  {journal} {Phys.
  Rev. B}\ }\textbf {\bibinfo {volume} {97}},\ \bibinfo {pages} {115115}
  (\bibinfo {year} {2018})}\BibitemShut {NoStop}%
\bibitem [{\citenamefont {Faure}\ \emph {et~al.}(2013)\citenamefont {Faure},
  \citenamefont {Mauchain}, \citenamefont {Papalazarou}, \citenamefont {Marsi},
  \citenamefont {Boschetto}, \citenamefont {Timrov}, \citenamefont {Vast},
  \citenamefont {Ohtsubo}, \citenamefont {Arnaud},\ and\ \citenamefont
  {Perfetti}}]{Faure2013}%
  \BibitemOpen
  \bibfield  {author} {\bibinfo {author} {\bibfnamefont {J.}~\bibnamefont
  {Faure}}, \bibinfo {author} {\bibfnamefont {J.}~\bibnamefont {Mauchain}},
  \bibinfo {author} {\bibfnamefont {E.}~\bibnamefont {Papalazarou}}, \bibinfo
  {author} {\bibfnamefont {M.}~\bibnamefont {Marsi}}, \bibinfo {author}
  {\bibfnamefont {D.}~\bibnamefont {Boschetto}}, \bibinfo {author}
  {\bibfnamefont {I.}~\bibnamefont {Timrov}}, \bibinfo {author} {\bibfnamefont
  {N.}~\bibnamefont {Vast}}, \bibinfo {author} {\bibfnamefont {Y.}~\bibnamefont
  {Ohtsubo}}, \bibinfo {author} {\bibfnamefont {B.}~\bibnamefont {Arnaud}}, \
  and\ \bibinfo {author} {\bibfnamefont {L.}~\bibnamefont {Perfetti}},\ }\href
  {\doibase 10.1103/PhysRevB.88.075120} {\bibfield  {journal} {\bibinfo
  {journal} {Phys. Rev. B}\ }\textbf {\bibinfo {volume} {88}},\ \bibinfo
  {pages} {075120} (\bibinfo {year} {2013})}\BibitemShut {NoStop}%
\bibitem [{\citenamefont {Klein}\ \emph {et~al.}(2018)\citenamefont {Klein},
  \citenamefont {Casula}, \citenamefont {Santos-Cottin}, \citenamefont
  {Audouard}, \citenamefont {Vignolles}, \citenamefont {F{\`{e}}ve},
  \citenamefont {Freulon}, \citenamefont {Pla{\c{c}}ais}, \citenamefont
  {Verseils}, \citenamefont {Yang}, \citenamefont {Paulatto},\ and\
  \citenamefont {Gauzzi}}]{Klein2018a}%
  \BibitemOpen
  \bibfield  {author} {\bibinfo {author} {\bibfnamefont {Y.}~\bibnamefont
  {Klein}}, \bibinfo {author} {\bibfnamefont {M.}~\bibnamefont {Casula}},
  \bibinfo {author} {\bibfnamefont {D.}~\bibnamefont {Santos-Cottin}}, \bibinfo
  {author} {\bibfnamefont {A.}~\bibnamefont {Audouard}}, \bibinfo {author}
  {\bibfnamefont {D.}~\bibnamefont {Vignolles}}, \bibinfo {author}
  {\bibfnamefont {G.}~\bibnamefont {F{\`{e}}ve}}, \bibinfo {author}
  {\bibfnamefont {V.}~\bibnamefont {Freulon}}, \bibinfo {author} {\bibfnamefont
  {B.}~\bibnamefont {Pla{\c{c}}ais}}, \bibinfo {author} {\bibfnamefont
  {M.}~\bibnamefont {Verseils}}, \bibinfo {author} {\bibfnamefont
  {H.}~\bibnamefont {Yang}}, \bibinfo {author} {\bibfnamefont {L.}~\bibnamefont
  {Paulatto}}, \ and\ \bibinfo {author} {\bibfnamefont {A.}~\bibnamefont
  {Gauzzi}},\ }\href {\doibase 10.1103/PhysRevB.97.075140} {\bibfield
  {journal} {\bibinfo  {journal} {Phys. Rev. B}\ }\textbf {\bibinfo {volume}
  {97}},\ \bibinfo {pages} {075140} (\bibinfo {year} {2018})}\BibitemShut
  {NoStop}%
\bibitem [{\citenamefont {Heyd}\ \emph {et~al.}(2003)\citenamefont {Heyd},
  \citenamefont {Scuseria},\ and\ \citenamefont {Ernzerhof}}]{heyd2003hybrid}%
  \BibitemOpen
  \bibfield  {author} {\bibinfo {author} {\bibfnamefont {J.}~\bibnamefont
  {Heyd}}, \bibinfo {author} {\bibfnamefont {G.~E.}\ \bibnamefont {Scuseria}},
  \ and\ \bibinfo {author} {\bibfnamefont {M.}~\bibnamefont {Ernzerhof}},\
  }\href@noop {} {\bibfield  {journal} {\bibinfo  {journal} {J. chem. Phys.}\
  }\textbf {\bibinfo {volume} {118}},\ \bibinfo {pages} {8207} (\bibinfo {year}
  {2003})}\BibitemShut {NoStop}%
\bibitem [{\citenamefont {Heyd}\ \emph {et~al.}(2006)\citenamefont {Heyd},
  \citenamefont {Scuseria},\ and\ \citenamefont {Ernzerhof}}]{ge2006erratum}%
  \BibitemOpen
  \bibfield  {author} {\bibinfo {author} {\bibfnamefont {J.}~\bibnamefont
  {Heyd}}, \bibinfo {author} {\bibfnamefont {G.~E.}\ \bibnamefont {Scuseria}},
  \ and\ \bibinfo {author} {\bibfnamefont {M.}~\bibnamefont {Ernzerhof}},\
  }\href@noop {} {\bibfield  {journal} {\bibinfo  {journal} {J. Chem. Phys}\
  }\textbf {\bibinfo {volume} {124}},\ \bibinfo {pages} {219906} (\bibinfo
  {year} {2006})}\BibitemShut {NoStop}%
\bibitem [{\citenamefont {Krukau}\ \emph {et~al.}(2006)\citenamefont {Krukau},
  \citenamefont {Vydrov}, \citenamefont {Izmaylov},\ and\ \citenamefont
  {Scuseria}}]{krukau2006influence}%
  \BibitemOpen
  \bibfield  {author} {\bibinfo {author} {\bibfnamefont {A.~V.}\ \bibnamefont
  {Krukau}}, \bibinfo {author} {\bibfnamefont {O.~A.}\ \bibnamefont {Vydrov}},
  \bibinfo {author} {\bibfnamefont {A.~F.}\ \bibnamefont {Izmaylov}}, \ and\
  \bibinfo {author} {\bibfnamefont {G.~E.}\ \bibnamefont {Scuseria}},\
  }\href@noop {} {\bibfield  {journal} {\bibinfo  {journal} {J. Chem. Phys.}\
  }\textbf {\bibinfo {volume} {125}},\ \bibinfo {pages} {224106} (\bibinfo
  {year} {2006})}\BibitemShut {NoStop}%
\bibitem [{\citenamefont {Perdew}\ \emph {et~al.}(1996)\citenamefont {Perdew},
  \citenamefont {Burke},\ and\ \citenamefont
  {Ernzerhof}}]{perdew1996generalized}%
  \BibitemOpen
  \bibfield  {author} {\bibinfo {author} {\bibfnamefont {J.~P.}\ \bibnamefont
  {Perdew}}, \bibinfo {author} {\bibfnamefont {K.}~\bibnamefont {Burke}}, \
  and\ \bibinfo {author} {\bibfnamefont {M.}~\bibnamefont {Ernzerhof}},\
  }\href@noop {} {\bibfield  {journal} {\bibinfo  {journal} {Phys. Rev. Lett.}\
  }\textbf {\bibinfo {volume} {77}},\ \bibinfo {pages} {3865} (\bibinfo {year}
  {1996})}\BibitemShut {NoStop}%
\bibitem [{\citenamefont {Perdew}\ and\ \citenamefont
  {Ernzerhof}(1996)}]{perdew1996phys}%
  \BibitemOpen
  \bibfield  {author} {\bibinfo {author} {\bibfnamefont {K.}~\bibnamefont
  {Perdew}, \bibfnamefont {J.~P.~Burke}}\ and\ \bibinfo {author} {\bibfnamefont
  {M.}~\bibnamefont {Ernzerhof}},\ }\href@noop {} {\bibfield  {journal}
  {\bibinfo  {journal} {Phys. Rev. Lett.}\ }\textbf {\bibinfo {volume} {78}},\
  \bibinfo {pages} {1396} (\bibinfo {year} {1996})}\BibitemShut {NoStop}%
\bibitem [{\citenamefont {Volkov}\ \emph {et~al.}(2019)\citenamefont {Volkov},
  \citenamefont {Sato}, \citenamefont {Schlaepfer}, \citenamefont {Kasmi},
  \citenamefont {Hartmann}, \citenamefont {Lucchini}, \citenamefont {Gallmann},
  \citenamefont {Rubio},\ and\ \citenamefont {Keller}}]{volkov2019attosecond}%
  \BibitemOpen
  \bibfield  {author} {\bibinfo {author} {\bibfnamefont {M.}~\bibnamefont
  {Volkov}}, \bibinfo {author} {\bibfnamefont {S.~A.}\ \bibnamefont {Sato}},
  \bibinfo {author} {\bibfnamefont {F.}~\bibnamefont {Schlaepfer}}, \bibinfo
  {author} {\bibfnamefont {L.}~\bibnamefont {Kasmi}}, \bibinfo {author}
  {\bibfnamefont {N.}~\bibnamefont {Hartmann}}, \bibinfo {author}
  {\bibfnamefont {M.}~\bibnamefont {Lucchini}}, \bibinfo {author}
  {\bibfnamefont {L.}~\bibnamefont {Gallmann}}, \bibinfo {author}
  {\bibfnamefont {A.}~\bibnamefont {Rubio}}, \ and\ \bibinfo {author}
  {\bibfnamefont {U.}~\bibnamefont {Keller}},\ }\href@noop {} {\bibfield
  {journal} {\bibinfo  {journal} {Nat. Phys.}\ }\textbf {\bibinfo {volume}
  {15}},\ \bibinfo {pages} {1145} (\bibinfo {year} {2019})}\BibitemShut
  {NoStop}%
\bibitem [{\citenamefont {Tancogne-Dejean}\ \emph {et~al.}(2018)\citenamefont
  {Tancogne-Dejean}, \citenamefont {Sentef},\ and\ \citenamefont
  {Rubio}}]{tancogne2018ultrafast}%
  \BibitemOpen
  \bibfield  {author} {\bibinfo {author} {\bibfnamefont {N.}~\bibnamefont
  {Tancogne-Dejean}}, \bibinfo {author} {\bibfnamefont {M.~A.}\ \bibnamefont
  {Sentef}}, \ and\ \bibinfo {author} {\bibfnamefont {A.}~\bibnamefont
  {Rubio}},\ }\href@noop {} {\bibfield  {journal} {\bibinfo  {journal} {Phys.
  Rev. Lett.}\ }\textbf {\bibinfo {volume} {121}},\ \bibinfo {pages} {097402}
  (\bibinfo {year} {2018})}\BibitemShut {NoStop}%
\bibitem [{\citenamefont {Gole{\v{z}}}\ \emph {et~al.}(2019)\citenamefont
  {Gole{\v{z}}}, \citenamefont {Boehnke}, \citenamefont {Eckstein},\ and\
  \citenamefont {Werner}}]{golevz2019dynamics}%
  \BibitemOpen
  \bibfield  {author} {\bibinfo {author} {\bibfnamefont {D.}~\bibnamefont
  {Gole{\v{z}}}}, \bibinfo {author} {\bibfnamefont {L.}~\bibnamefont
  {Boehnke}}, \bibinfo {author} {\bibfnamefont {M.}~\bibnamefont {Eckstein}}, \
  and\ \bibinfo {author} {\bibfnamefont {P.}~\bibnamefont {Werner}},\
  }\href@noop {} {\bibfield  {journal} {\bibinfo  {journal} {Phys. Rev. B}\
  }\textbf {\bibinfo {volume} {100}},\ \bibinfo {pages} {041111} (\bibinfo
  {year} {2019})}\BibitemShut {NoStop}%
\bibitem [{\citenamefont {Golez}\ \emph {et~al.}(2019)\citenamefont {Golez},
  \citenamefont {Eckstein},\ and\ \citenamefont {Werner}}]{golez2019multi}%
  \BibitemOpen
  \bibfield  {author} {\bibinfo {author} {\bibfnamefont {D.}~\bibnamefont
  {Golez}}, \bibinfo {author} {\bibfnamefont {M.}~\bibnamefont {Eckstein}}, \
  and\ \bibinfo {author} {\bibfnamefont {P.}~\bibnamefont {Werner}},\
  }\href@noop {} {\bibfield  {journal} {\bibinfo  {journal} {arXiv preprint
  arXiv:1903.08713}\ } (\bibinfo {year} {2019})}\BibitemShut {NoStop}%
\bibitem [{\citenamefont {Ayral}\ \emph {et~al.}(2017)\citenamefont {Ayral},
  \citenamefont {Biermann}, \citenamefont {Werner},\ and\ \citenamefont
  {Boehnke}}]{ayral2017influence}%
  \BibitemOpen
  \bibfield  {author} {\bibinfo {author} {\bibfnamefont {T.}~\bibnamefont
  {Ayral}}, \bibinfo {author} {\bibfnamefont {S.}~\bibnamefont {Biermann}},
  \bibinfo {author} {\bibfnamefont {P.}~\bibnamefont {Werner}}, \ and\ \bibinfo
  {author} {\bibfnamefont {L.}~\bibnamefont {Boehnke}},\ }\href@noop {}
  {\bibfield  {journal} {\bibinfo  {journal} {Phys. Rev. B}\ }\textbf {\bibinfo
  {volume} {95}},\ \bibinfo {pages} {245130} (\bibinfo {year}
  {2017})}\BibitemShut {NoStop}%
\bibitem [{\citenamefont {Giannozzi}\ \emph {et~al.}(2009)\citenamefont
  {Giannozzi}, \citenamefont {Baroni}, \citenamefont {Bonini}, \citenamefont
  {Calandra}, \citenamefont {Car}, \citenamefont {Cavazzoni}, \citenamefont
  {Ceresoli}, \citenamefont {Chiarotti}, \citenamefont {Cococcioni},
  \citenamefont {Dabo}, \citenamefont {Corso}, \citenamefont {de~Gironcoli},
  \citenamefont {Fabris}, \citenamefont {Fratesi}, \citenamefont {Gebauer},
  \citenamefont {Gerstmann}, \citenamefont {Gougoussis}, \citenamefont
  {Kokalj}, \citenamefont {Lazzeri}, \citenamefont {Martin-Samos},
  \citenamefont {Marzari}, \citenamefont {Mauri}, \citenamefont {Mazzarello},
  \citenamefont {Paolini}, \citenamefont {Pasquarello}, \citenamefont
  {Paulatto}, \citenamefont {Sbraccia}, \citenamefont {Scandolo}, \citenamefont
  {Sclauzero}, \citenamefont {Seitsonen}, \citenamefont {Smogunov},
  \citenamefont {Umari},\ and\ \citenamefont {Wentzcovitch}}]{QE-2009}%
  \BibitemOpen
  \bibfield  {author} {\bibinfo {author} {\bibfnamefont {P.}~\bibnamefont
  {Giannozzi}}, \bibinfo {author} {\bibfnamefont {S.}~\bibnamefont {Baroni}},
  \bibinfo {author} {\bibfnamefont {N.}~\bibnamefont {Bonini}}, \bibinfo
  {author} {\bibfnamefont {M.}~\bibnamefont {Calandra}}, \bibinfo {author}
  {\bibfnamefont {R.}~\bibnamefont {Car}}, \bibinfo {author} {\bibfnamefont
  {C.}~\bibnamefont {Cavazzoni}}, \bibinfo {author} {\bibfnamefont
  {D.}~\bibnamefont {Ceresoli}}, \bibinfo {author} {\bibfnamefont {G.~L.}\
  \bibnamefont {Chiarotti}}, \bibinfo {author} {\bibfnamefont {M.}~\bibnamefont
  {Cococcioni}}, \bibinfo {author} {\bibfnamefont {I.}~\bibnamefont {Dabo}},
  \bibinfo {author} {\bibfnamefont {A.~D.}\ \bibnamefont {Corso}}, \bibinfo
  {author} {\bibfnamefont {S.}~\bibnamefont {de~Gironcoli}}, \bibinfo {author}
  {\bibfnamefont {S.}~\bibnamefont {Fabris}}, \bibinfo {author} {\bibfnamefont
  {G.}~\bibnamefont {Fratesi}}, \bibinfo {author} {\bibfnamefont
  {R.}~\bibnamefont {Gebauer}}, \bibinfo {author} {\bibfnamefont
  {U.}~\bibnamefont {Gerstmann}}, \bibinfo {author} {\bibfnamefont
  {C.}~\bibnamefont {Gougoussis}}, \bibinfo {author} {\bibfnamefont
  {A.}~\bibnamefont {Kokalj}}, \bibinfo {author} {\bibfnamefont
  {M.}~\bibnamefont {Lazzeri}}, \bibinfo {author} {\bibfnamefont
  {L.}~\bibnamefont {Martin-Samos}}, \bibinfo {author} {\bibfnamefont
  {N.}~\bibnamefont {Marzari}}, \bibinfo {author} {\bibfnamefont
  {F.}~\bibnamefont {Mauri}}, \bibinfo {author} {\bibfnamefont
  {R.}~\bibnamefont {Mazzarello}}, \bibinfo {author} {\bibfnamefont
  {S.}~\bibnamefont {Paolini}}, \bibinfo {author} {\bibfnamefont
  {A.}~\bibnamefont {Pasquarello}}, \bibinfo {author} {\bibfnamefont
  {L.}~\bibnamefont {Paulatto}}, \bibinfo {author} {\bibfnamefont
  {C.}~\bibnamefont {Sbraccia}}, \bibinfo {author} {\bibfnamefont
  {S.}~\bibnamefont {Scandolo}}, \bibinfo {author} {\bibfnamefont
  {G.}~\bibnamefont {Sclauzero}}, \bibinfo {author} {\bibfnamefont {A.~P.}\
  \bibnamefont {Seitsonen}}, \bibinfo {author} {\bibfnamefont {A.}~\bibnamefont
  {Smogunov}}, \bibinfo {author} {\bibfnamefont {P.}~\bibnamefont {Umari}}, \
  and\ \bibinfo {author} {\bibfnamefont {R.~M.}\ \bibnamefont {Wentzcovitch}},\
  }\href@noop {} {\bibfield  {journal} {\bibinfo  {journal} {J. Phys: Condens.
  Mat.}\ }\textbf {\bibinfo {volume} {21}},\ \bibinfo {pages} {395502}
  (\bibinfo {year} {2009})}\BibitemShut {NoStop}%
\bibitem [{\citenamefont {Giannozzi}\ \emph {et~al.}(2017)\citenamefont
  {Giannozzi}, \citenamefont {Andreussi}, \citenamefont {Brumme}, \citenamefont
  {Bunau}, \citenamefont {Nardelli}, \citenamefont {Calandra}, \citenamefont
  {Car}, \citenamefont {Cavazzoni}, \citenamefont {Ceresoli}, \citenamefont
  {Cococcioni} \emph {et~al.}}]{giannozzi2017}%
  \BibitemOpen
  \bibfield  {author} {\bibinfo {author} {\bibfnamefont {P.}~\bibnamefont
  {Giannozzi}}, \bibinfo {author} {\bibfnamefont {O.}~\bibnamefont
  {Andreussi}}, \bibinfo {author} {\bibfnamefont {T.}~\bibnamefont {Brumme}},
  \bibinfo {author} {\bibfnamefont {O.}~\bibnamefont {Bunau}}, \bibinfo
  {author} {\bibfnamefont {M.~B.}\ \bibnamefont {Nardelli}}, \bibinfo {author}
  {\bibfnamefont {M.}~\bibnamefont {Calandra}}, \bibinfo {author}
  {\bibfnamefont {R.}~\bibnamefont {Car}}, \bibinfo {author} {\bibfnamefont
  {C.}~\bibnamefont {Cavazzoni}}, \bibinfo {author} {\bibfnamefont
  {D.}~\bibnamefont {Ceresoli}}, \bibinfo {author} {\bibfnamefont
  {M.}~\bibnamefont {Cococcioni}},  \emph {et~al.},\ }\href@noop {} {\bibfield
  {journal} {\bibinfo  {journal} {J. Phys: Condens. Mat.}\ }\textbf {\bibinfo
  {volume} {29}},\ \bibinfo {pages} {465901} (\bibinfo {year}
  {2017})}\BibitemShut {NoStop}%
\bibitem [{\citenamefont {Grey}\ and\ \citenamefont
  {Steinfink}(1970)}]{Grey1970}%
  \BibitemOpen
  \bibfield  {author} {\bibinfo {author} {\bibfnamefont {I.~E.}\ \bibnamefont
  {Grey}}\ and\ \bibinfo {author} {\bibfnamefont {H.}~\bibnamefont
  {Steinfink}},\ }\href {\doibase 10.1021/ja00720a015} {\bibfield  {journal}
  {\bibinfo  {journal} {J. Am. Chem. Soc.}\ }\textbf {\bibinfo {volume} {92}},\
  \bibinfo {pages} {5093} (\bibinfo {year} {1970})}\BibitemShut {NoStop}%
\end{thebibliography}%

\end{document}